\documentclass[reprint,amsmath,amssymb,aps,superscriptaddress,pre]{revtex4-2}
\usepackage{graphicx}% Include figure files
\usepackage{color}
\usepackage{hyperref}% add hypertext capabilities
\usepackage{longtable}
\usepackage{booktabs}
\usepackage{siunitx}
\usepackage{bm}
\usepackage{dcolumn}% Align table columns on decimal point
\usepackage{nicefrac}

\renewcommand{\eqref}[1]{Eq.~(\ref{#1})}
\newcommand{\eqsref}[1]{Eqs.~(\ref{#1})}
\newcommand{\figref}[1]{Fig.~\ref{#1}}

\newcommand{\BI}{{\text{B}\rightarrow\text{I}}}
\newcommand{\IB}{{\text{I}\rightarrow\text{B}}}
\newcommand{\kIB}{k_\IB}
\newcommand{\kBI}{k_\BI}
\newcommand{\jB}{j_\text{B}}
\newcommand{\jI}{j_\text{I}}
\newcommand{\jBI}{j_\text{B/I}}
\newcommand{\JB}{J_\text{B}}

\newcommand{\sBI}{s_\BI}
\newcommand{\sIB}{s_\IB}
\newcommand{\rhov}{\rho^\text{v}}
\newcommand{\rhovf}{\rho^\text{v,flat}}
\newcommand{\rhoB}{\rho_\text{B}}
\newcommand{\rhoI}{\rho_\text{I}}
\newcommand{\rhoBI}{\rho_\text{B/I}}
\newcommand{\rhoBv}{\rho_\text{B}^\text{v}}
\newcommand{\rhoIv}{\rho_\text{I}^\text{v}}

\newcommand{\pB}{p_\text{B}}
\newcommand{\pI}{p_\text{I}}
\newcommand{\pBI}{p_\text{B/I}}
\newcommand{\Dmu}{\Delta\mu}
\newcommand{\Df}{\Delta f}
\newcommand{\Dfeff}{\Delta f_{\text{eff}}}
\newcommand{\Dh}{\Delta h}

\newcommand{\nB}{n_\text{B}}

\begin{document}

\title{Interfacial Effects Determine Nonequilibrium Phase Behaviors\\in Chemically Driven Fluids}

\author{Yongick Cho}
\affiliation{Department of Chemistry, Princeton University, Princeton, NJ 08544, USA}
\author{William M. Jacobs}
\email{wjacobs@princeton.edu}
\affiliation{Department of Chemistry, Princeton University, Princeton, NJ 08544, USA}

\begin{abstract}
Coupling between chemical fuel consumption and phase separation can lead to condensation at a nonequilibrium steady state, resulting in phase behaviors that are not described by equilibrium thermodynamics.
Theoretical models of such ``chemically driven fluids'' typically invoke near-equilibrium approximations at small length scales.
However, because dissipation occurs due to both molecular-scale chemical reactions and mesoscale diffusive transport, it has remained unclear which properties of phase-separated reaction--diffusion systems can be assumed to be at an effective equilibrium.
Here we use microscopic simulations to show that mesoscopic fluxes are dependent on nonequilibrium fluctuations at phase-separated interfaces.
We further develop a first-principles theory to predict nonequilibrium coexistence curves, localization of mesoscopic fluxes near phase-separated interfaces, and droplet size-scaling relations in good agreement with simulations.
Our findings highlight the central role of interfacial properties in governing nonequilibrium condensation and have broad implications for droplet nucleation, coarsening, and size control in chemically driven fluids.
\end{abstract}

\date{\today}

\maketitle

Although phase-separated biomolecular condensates are commonly modeled within the framework of equilibrium thermodynamics~\cite{lyon2021review,villegas2022review,jacobs2023review,pappu2023review}, nonequilibrium chemical reactions can strongly affect the phase behavior and dynamics of intracellular biomolecular mixtures~\cite{soding2020mechanisms,Karsten2022DEADbox}.
For example, ATPase ParA controls the subcellular localization of ParB condensates at specific DNA sequences by opposing coarsening~\cite{Guilhas2020ATP}.
By contrast, kinase DivJ maintains its activity under nutrient scarcity by partitioning into condensates whose formation is promoted by low ATP levels~\cite{Saurabh2022ATP}.
ATP-dependent reactions can also tune the nucleation rates of biomolecular condensates, as observed in the formation of LAT clusters in the course of T-cell antigen discrimination~\cite{McAfee2022LAT}.
Controlling such chemical reactions allows cells to regulate condensate assembly and dissolution independently of protein and RNA expression levels, taking advantage of the fact that phase-separated condensates are not permanent structures like traditional membrane-bound organelles, and thus enabling rapid spatial reorganization of intracellular biochemistry in response to stimuli~\cite{sanders2020competing,Yang2020G3BP1,Guillen-Boixet2020G3BP1,shimobayashi2021nucleation,strom2024interplay}.

Predicting the consequences of coupled chemical reactions and phase separation from molecular-scale details remains an outstanding theoretical challenge.
When a chemical fuel such as ATP is continually available at constant chemical potential, the free-energy difference between chemical fuel and waste can be used to drive conformational transitions or chemical modifications of macromolecules, so that the populations of these macromolecular states are maintained out of equilibrium in a nonequilibrium steady state.
If these macromolecular states have different tendencies to condense, then the fuel-dependent reactions and phase separation become coupled~\cite{glotzer1995reaction,berry2018physical,weber2019physics,Julicher2024Droplet}, potentially leading to arrested coarsening~\cite{zwicker2015suppression,wurtz2018chemical,Bauermann2024extensive}, self-induced diffusiophoresis~\cite{Nasouri2020phoresis,Demarchi2023propulsion,Hafner2024diffusiophoresis,Kriebisch2024activetransport}, and steady-state nonequilibrium condensate morphologies~\cite{Donau2022coacervate,Bergmann2023Spherical,Prathyusha2023Structure}.
However, the tools of equilibrium statistical mechanics that are used to predict equilibrium phase diagrams, condensate morphologies, and phase-transformation kinetics from molecular-scale details do not apply to nonequilibrium steady states.
There is thus a need for new approaches to connect ``dissipative'' reactions~\cite{grzybowski2005micro,fialkowski2006principles,riess2020design,penocchio2019thermodynamic,avanzini2024nonequilibrium}, which take place at the level of individual macromolecules, to their collective effects in phase-separated mixtures, where steady-state fluxes can appear on mesoscopic scales.

These challenges are particularly acute at condensate interfaces, where phases with potentially different dissipation rates come into contact.
For example, dissipative reaction rates on either side of a condensate interface can be modified by fuel-dependent enzymes that partition into or out of specific condensates~\cite{hondele2020membraneless,oflynn2021role}.
Theoretical analyses of mesoscale fluxes in these systems typically assume that phase-separated interfaces can be described by equilibrium or near-equilibrium models~\cite{berry2018physical,weber2019physics,Julicher2024Droplet,kirschbaum2021controlling,zwicker2022intertwined,bauermann2022energy}.
Yet this assumption contrasts with simulation results showing that dissipative chemical reactions can modify microscopic fluctuations and lead to nonequilibrium interfacial properties~\cite{cho2023nucleation,cho2023interface}, as has also been seen in other classes of nonequilibrium phase-separated fluids~\cite{redner2016classical,paliwal2017non,fausti2021capillary}.
Such interfacial effects may have significant implications for regulating condensate assembly in living cells due to the dominant role that interfacial properties play in the kinetics of droplet nucleation, growth, and coarsening.
However, the relationship between microscopic interfacial fluctuations and the mesoscopic fluxes that emerge in phase-separated chemically driven fluids has not been established.

\begin{figure*}
    \centering
    \includegraphics[width=0.725\textwidth]{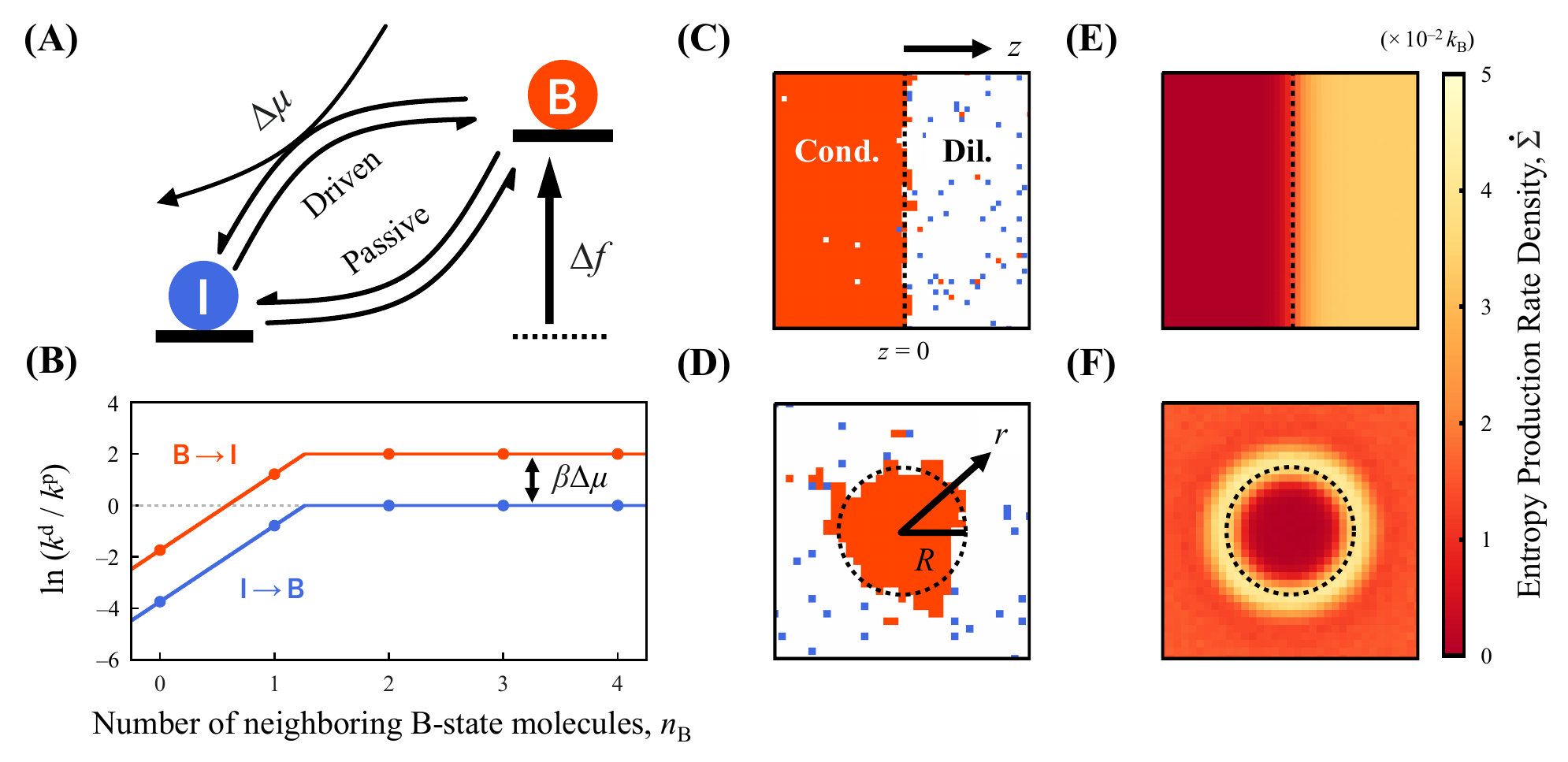}
    \caption{\textbf{A molecular-scale model of nonequilibrium phase coexistence in chemically driven fluids.}
    (A) Schematic of passive and driven reaction pathways between the bonding, B, and inert, I, molecular states with internal energy difference $\Df$. Only the driven reaction pathway is coupled to the nonequilibrium chemical drive, $\Dmu$.
    (B) In our chemical reaction scheme, the driven reaction rate, $k^\text{d}$, is enhanced over the passive rate, $k^\text{p}$, in both the $\BI$ and $\IB$ directions at high local B-state concentrations, as determined from the number of neighboring B-state molecules, $\nB$.
    (C,D) Example configurations of coexisting condensed and dilute phases with (C) macroscopically flat and (D) curved interfaces at a nonequilibrium steady state. The distance from the average position of the interface, $z$, and from the center of mass of the droplet, $r$, are indicated.
    (E,F) The time-averaged steady-state entropy production rate density, $\dot{\Sigma}$, for the (E) flat and (F) curved interfaces.
    In panels (C--F), the black dotted lines show the Gibbs dividing surfaces, which indicate the average locations of the flat and curved phase-separated interfaces at $z = 0$ and $r = R$, respectively.
    See \figref{fig:figure2} for simulation parameters.
    }
    \label{fig:figure1}
\end{figure*}

In this work, we show that steady-state mesoscopic fluxes are directly related to nonequilibrium fluctuations in phase-separated chemically driven fluids, and we propose a first-principles theory to explain how these microscopic and mesoscopic effects are coupled at condensate interfaces.
Our simulations and theoretical analysis demonstrate that mesoscopic fluxes originate from, and are thus concentrated near, condensate interfaces despite the presence of molecular-level dissipation throughout the system.
We further show that both the conditions for phase coexistence and the size scaling of condensates are strongly influenced by nonequilibrium interfacial effects that can be understood via a microscopic theory.
Our theory predicts a broad range of conditions in which nonequilibrium interfacial effects make significant contributions to the stability and dynamics of nonequilibrium condensates, highlighting the importance of condensate interfaces in chemically driven phase-separated mixtures.

\section*{Microscopic Model of a Chemically Driven Nonequilibrium Steady State}

As a minimal model of a chemically driven fluid, we consider a ternary system in which macromolecules (referred to as molecules from this point on) undergo chemical reactions between conformational states while diffusing in an implicit solvent.
Each molecule is in either a bonding (B) state or an inert (I) state, whose internal energy differs by $\Df$ (\figref{fig:figure1}A).
Importantly, molecules in the B state experience a short-ranged attractive interaction with other B-state molecules only.
The size of a molecule sets the unit length scale, and unoccupied space represents the solvent.

We model the fluid using a two-dimensional square lattice, in which the dimensionless nearest-neighbor interaction energy among B-state molecules is $\beta\epsilon = -2.95$ and $\beta$ is the inverse temperature of the solvent.
This choice ensures that the fluid can phase-separate into dilute and condensed phases at equilibrium \cite{pathria1996statistical}.
B-state molecules engage in intermolecular interactions according to the potential energy $u = \epsilon\nB$, where $\nB$ is the number of nearest-neighbor B-state molecules.
On the other hand, I-state molecules always have $u = 0$, as they do not participate in attractive intermolecular interactions.
Molecules diffuse with mobility $\Lambda$, so that the mean squared displacement over time is proportional to $\Lambda$ in the dilute limit.
In simulations, diffusion is affected by both the attractive interactions among B-state molecules and the hard-core repulsion that prohibits the occupancy of more than one molecule per lattice site.
The rate of diffusion to an adjacent lattice site is thus $\Lambda\exp(\beta u)\delta_{\text{unocc}}/4$, where $\delta_{\text{unocc}}$ is 1 if the destination lattice site is unoccupied and 0 otherwise, and the factor of 4 accounts for the coordination number of the square lattice (see \textit{Materials and Methods}).

Chemical reactions between B and I states proceed through two distinct reaction pathways.
For simplicity, we consider a single driven pathway that is coupled to a chemical drive, $\Dmu$, which originates from the chemical potential difference between chemical fuel and waste.
Positive values of $\Dmu$ accelerate reactions in the $\text{B}\rightarrow\text{I}$ direction (\figref{fig:figure1}A).
By contrast, the passive pathway is independent of $\Dmu$.
We note that this scheme could also be modified to represent two driven pathways in opposing directions with different chemical drives.
Utilizing a reaction scheme from prior work~\cite{cho2023nucleation,cho2023interface}, we set the $\text{I}\rightarrow\text{B}$ reaction rate constants as
\begin{equation}\label{eq:forward_rates}
\begin{aligned}
    \kIB^\text{p} &= 1 \\
    \kIB^\text{d} &= \eta \min[1,\exp(-\beta\epsilon\nB - \beta\Df - \beta\Dmu)],
\end{aligned}
\end{equation}
where the superscripts $\text{p}$ and $\text{d}$ indicate the passive and driven pathways, respectively, and $\eta$ is a constant that controls the ratio of reactive fluxes along the two pathways.
The passive rate is a constant that sets the fundamental time unit for the simulation, whereas the $\nB$-dependence of the driven reaction rate accounts for concentration-dependent rate constants.
Specifically, the driven reaction rate constant in \eqref{eq:forward_rates} is enhanced at high B-state concentrations (\figref{fig:figure1}B).
This particular choice can describe, for example, a scenario in which the enzyme that catalyzes the driven pathway partitions into the B-state-rich condensate~\cite{kirschbaum2021controlling} or, alternatively, a scenario in which B-state molecules directly catalyze reactions involving other nearby molecules subject to a maximum turnover rate.
Alternative schemes, which confirm the generality of our results, are presented in the \textit{Supplementary Information (SI)}.
We fix $\eta = 1$ throughout this paper because the mesoscopic effects of the driven reactions are maximized when the reactive fluxes along the two pathways are comparable~\cite{cho2023interface} (Fig.~S1).

Assuming that the solvent is at equilibrium, stochastic thermodynamics~\cite{van2015ensemble} constrains the rate constants along the driven and passive pathways in terms of the total free-energy change:
\begin{equation}\label{eq:local_detailed_balance}
\begin{aligned}
    \kBI^\text{p} / \kIB^\text{p} &= \exp(\beta\Df + \beta \epsilon\nB) \\
    \kBI^\text{d} / \kIB^\text{d} &= \exp(\beta\Df + \beta \epsilon\nB + \beta\Dmu).
\end{aligned}
\end{equation}
The driven reaction in the $\text{B}\rightarrow\text{I}$ direction is thus accelerated over the $\text{I}\rightarrow\text{B}$ direction by a factor $\exp(\beta\Dmu)$, regardless of the local B-state concentration (\figref{fig:figure1}B).
Because the overall chemical reaction rate is the sum of the rates along the two pathways, applying nonzero $\Dmu$ means that the system is no longer governed by equilibrium statistical mechanics.
The probability of observing a specific sequence of reactions and transitions of molecules between lattice sites is therefore different from that of the time-reversed sequence of these events, leading to steady-state entropy production~\cite{van2015ensemble}.
Importantly, entropy production is a direct measure of molecular-scale dissipation and is always positive for nonzero $\Dmu$.

Kinetic Monte Carlo simulations of a condensed-phase droplet with a curved interface of radius $R$ (\figref{fig:figure1}D) and a flat interface in the macroscopic limit $R \rightarrow\infty$ (\figref{fig:figure1}C) allow us to calculate the spatial dependence of the steady-state entropy production rate density (see \textit{Materials and Methods}).
The rate of entropy production is lower in the condensed phase due to the slower rate of $\text{B}\rightarrow\text{I}$ reactions when the attractive interactions among B-state molecules must be disrupted (\figref{fig:figure1}E,F).
In the dilute phase, entropy production is primarily due to ``futile cycles'', which execute closed loops around the driven and passive pathways.
However, we observe that the entropy production rate density does not necessarily decrease monotonically upon approaching the condensed phase, but may even increase again near the phase-separated interface (Fig.~S2).
This observation hints that nonequilibrium effects at the interface may differ from the bulk behavior of the coexisting phases.

\section*{Mesoscopic Fluxes are Concentrated near the Interface}

\begin{figure}
    \centering
    \includegraphics[width=\linewidth]{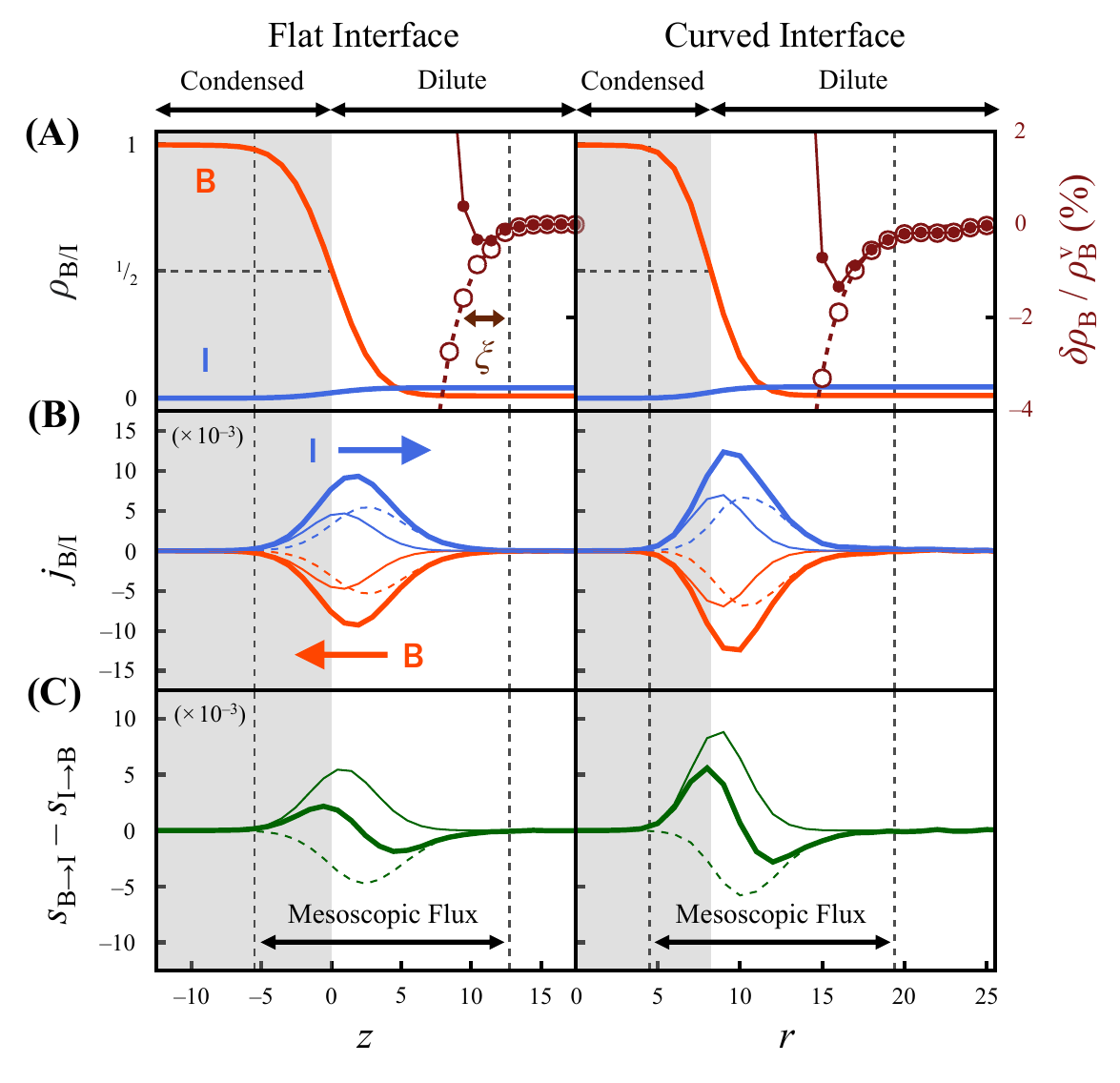}
    \caption{\textbf{Nonequilibrium diffusive and reactive fluxes are concentrated near condensate interfaces.}
    (A) \textit{Left axis:} Concentration profiles for B-state (orange) and I-state (blue) molecules as a function of the distance from the average location of the interface at $z = 0$ for the flat interface and as a function of the distance from the center of the droplet, $r$, for the curved interface located at $R = 8.46$ (cf.~\figref{fig:figure1}C,D).
    Shaded and unshaded regions indicate the condensed and dilute phases, respectively, separated by the Gibbs dividing surface.
    \textit{Right axis:} Zoomed-in plot of the change in the B-state concentration, $\delta\rhoB \equiv \rhoB - \rhoBv$, relative to the far-field steady-state value, $\rhoBv$ (brown filled circles).
    The contribution to this profile from the dilute phase (brown open circles) is shown separately since the contribution from the fluctuating interface obscures the steady-state concentration gradient in the dilute phase.
    Brown dashed lines show a fit to the theory with $\xi$ independently measured in the dilute phase.
    (B) Net diffusive flux densities, $\jBI$, and
    (C) the net reactive flux density in the $\BI$ direction, $\sBI-\sIB$.
    In both panels, thick solid lines indicate the total simulated flux densities, whereas thin solid and dotted lines are the contributions from the instantaneous interface and the dilute phase, respectively.
    The vertical dashed lines mark the boundaries of the mesoscopic flux region where $|\jB|$ is greater than 1\% of its peak value.
    Simulation parameters are $\beta\Dmu = 2$, $\Lambda = 10^2$, and $\beta\Df = 1.7337$, resulting in far-field total concentrations, $\rhov \equiv \rhoBv + \rhoIv$, of $0.05$ and $0.0548$ for the flat and curved interfaces, respectively.
    Note that the left and right columns correspond to \figref{fig:figure1}C,E and \figref{fig:figure1}D,F, respectively.
    }
    \label{fig:figure2}
\end{figure}

Although entropy production occurs throughout the entire system, we find that mesoscopic nonequilibrium effects, represented by the net diffusive and reactive fluxes, exist only near the interface.
Example simulations for both curved and flat interfaces are shown in \figref{fig:figure2}, where the time-averaged location of the interface is specified by the Gibbs dividing surface at which the B-state concentration is $\rhoB = \nicefrac{1}{2}$ (\figref{fig:figure2}A).
The net diffusive flux densities, $\jBI$ (\figref{fig:figure2}B), and the net reactive flux density in the $\BI$ direction, $\sBI-\sIB$ (\figref{fig:figure2}C), are nonzero only near the Gibbs dividing surface (see \textit{Materials and Methods}).
These mesoscopic fluxes decay rapidly to zero in the bulk phases in simulations with both flat and curved interfaces, primarily because the contributions from futile cycles average out.

To understand the origin of these mesoscopic effects, we divide the observed net fluxes (thick solid lines in \figref{fig:figure2}B,C) into contributions from molecules in the bulk dilute phase (thin solid lines) and from molecules that are within one lattice spacing of the instantaneous interface between the two phases (dashed lines; see \textit{Materials and Methods}).
The contribution from molecules in the condensed phase is negligible and is thus not shown for clarity.
For both the diffusive and reactive fluxes, the interfacial contribution is centered at the Gibbs dividing surface by construction.
We note that the time-averaged interfacial contribution to each of these fluxes, as well as the time-averaged concentration profiles of the B and I-state molecules in \figref{fig:figure2}A, is spatially smeared out due to steady-state fluctuations of the interface, even though the instantaneous interface is locally sharp (\figref{fig:figure1}C,D).
By contrast, the contribution from molecules in the dilute phase necessarily occurs to the right of the interface, so that the total mesoscopic fluxes are asymmetric with respect to the Gibbs dividing surface.
The dilute-phase contribution to the diffusive flux arises due to a steady-state concentration gradient, in which the B-state concentration monotonically increases towards the far-field value, $\rhoBv$, as a function of the distance from the interface.
This gradient can be observed directly by plotting the contribution to $\delta\rhoB \equiv \rhoB - \rhoBv$ from molecules in the bulk dilute phase, so that interfacial fluctuations do not obscure this effect (open circles in \figref{fig:figure2}A; see also Fig.~S3).
In sum, the total net diffusive flux is peaked on the dilute-phase side of the interface (\figref{fig:figure2}B), in alignment with the inflection point of the total net reactive flux (\figref{fig:figure2}C).

This analysis suggests a simple mechanistic picture of the mesoscopic fluxes, in which the molecules execute a reaction-diffusion cycle at steady state.
When $\Dmu > 0$, excess B-state molecules in the dilute phase, resulting from dilute-phase $\IB$ reactions, drive a flux of B-state molecules towards the condensed phase until they are adsorbed at the interface.
Adsorbed B-state molecules tend to undergo $\BI$ chemical reactions ($\sBI - \sIB > 0$) and subsequently desorb from the interface due to the lack of attractive interactions between I-state molecules and the condensed phase.
I-state molecules are eventually converted back to the B-state in the dilute phase ($\sBI - \sIB < 0$).
More generally, the direction of these fluxes may be reversed if either the sign of $\Dmu$ or the $\nB$-dependence of the reaction rate constants is reversed (Fig.~S4).

\section*{Interfacial Fluctuations Deviate from Thermal Equilibrium}

\begin{figure}
    \centering
    \includegraphics[width=\linewidth]{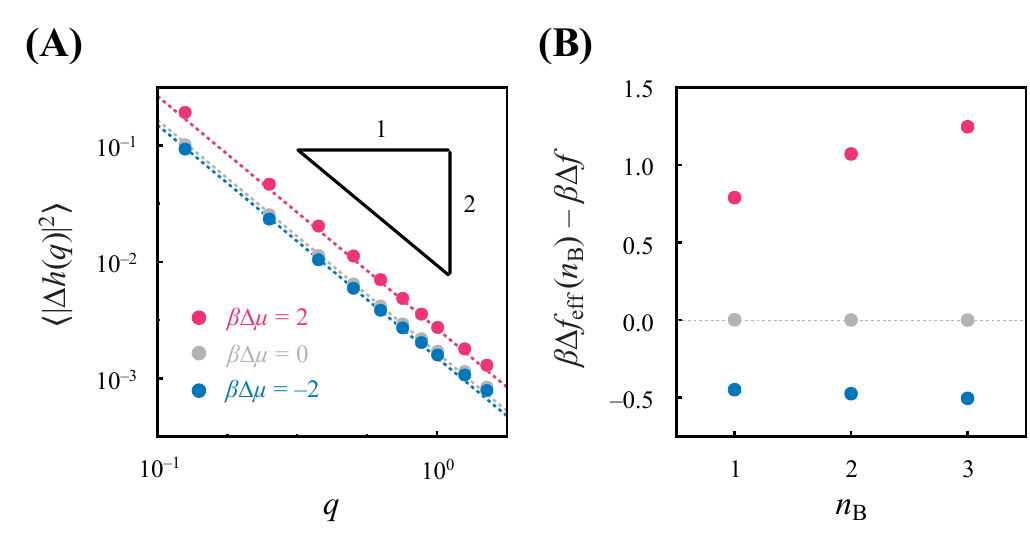}
    \caption{\textbf{Interfacial fluctuations can deviate from equilibrium in the presence of driven chemical reactions.}
      (A)~The steady-state power spectrum of height fluctuations of a flat interface, $\langle |\Dh(q)|^2 \rangle$, with respect to a wavevector parallel to the interface, $q$. Marks indicate the measured spectrum, and the dotted lines are a fit to the equipartition relation, $\langle |\Dh(q)|^2 \rangle \propto q^{-2}$, in the range $q\in[10^{-1},10^0]$.
      (B)~A measure of the non-Boltzmann distribution of B and I-state populations on the interface, $\Dfeff(\nB)$, compared to the internal energy difference, $\Df$.
      The simulation parameters are $(\beta\Df,\beta\Dmu) = (1.7337,2;\text{pink}), (2.8362,0;\text{gray})$, and $(3.3150,-2;\text{teal})$ at constant $\Lambda = 10^2$, which all result in a constant $\rhov=0.05$.
      Note that the $\beta\Dmu=2$ condition corresponds to the left column of \figref{fig:figure2}.}
    \label{fig:figure2.5}
\end{figure}

Examination of the microscopic configurations reveals further deviations from thermal equilibrium at the microscopic scale.
We first compute the spectrum of interfacial height fluctuations, commonly referred to as capillary waves~\cite{Rowlinson1981molecularcapillarity}, which describes spatial correlations in the position of the instantaneous interface relative to the Gibbs dividing surface.
At equilibrium, capillary fluctuations of a macroscopically flat interface (i.e., $R \rightarrow \infty$; cf.~\figref{fig:figure1}C) only depend on the reduced B-state interaction energy, $\beta\epsilon$, and the transverse dimension of the simulation, $L$, since the I-state molecules are energetically indistinguishable from the implicit solvent in our simulations.
Moreover, due to the equipartition theorem, the average power spectrum of capillary fluctuations, $\langle|\Dh(q)|^2\rangle$, is proportional to $1/\beta\sigma q^{2}$ in equilibrium, where $\sigma$ is the interfacial tension and $q$ is a wavevector parallel to the interface~\cite{Rowlinson1981molecularcapillarity}.
For the example conditions analyzed in \figref{fig:figure2}, we find that capillary fluctuations in the presence of steady-state fluxes deviate only slightly from the equipartition $q^{-2}$ scaling at small $q$ but differ significantly from the equilibrium spectrum in their overall amplitude (pink marks in \figref{fig:figure2.5}A).
This upshift in the spectrum can be interpreted as a reduction in the interfacial tension, $\sigma$, relative to the equilibrium value.
This effect also leads to a broadening of the time-averaged interfacial width (\figref{fig:figure2}A), which can be quantified by the mean-squared interfacial fluctuation, $\langle\Dh^2\rangle$.
Reversing the direction of the chemical drive under otherwise equivalent conditions results in an increased interfacial tension (teal marks in \figref{fig:figure2.5}A) and a narrowing of the time-averaged interfacial width, along with reversed mesoscopic fluxes (Fig.~S4).

Next, we look more closely at the identities of the molecules within one lattice spacing of the instantaneous interface.
At equilibrium, the ratio of B-state to I-state molecules follows the Boltzmann distribution, $\pB(\nB)/\pI(\nB) = \exp(-\beta\Df - \beta\epsilon\nB)$,
where $\nB \in \{1,2,3\}$ at the interface.
However, when steady-state fluxes are present, we find that the conditional distributions $p_\text{B/I}(\nB)$ are incompatible with Boltzmann statistics.
These deviations are most easily seen by inverting the Boltzmann distribution to define $\beta\Dfeff(\nB)\equiv -\ln[\pB(\nB)/\pI(\nB)]-\beta\epsilon\nB$~\cite{cho2023interface}.
For the example conditions analyzed in \figref{fig:figure2}, $\Dfeff$ increases with $\nB$ (pink marks in \figref{fig:figure2.5}B), indicating that the nonequilibrium population of molecules at the interface is biased more towards the B state when $\nB=1$ and more towards the I state when $\nB=3$.
An analogous calculation of the marginal distribution of lattice sites with $\nB$ B-state neighbors on the interface, $p(\nB)$, further reveals a decrease in the populations of $\nB=1$ and $\nB=3$ lattice sites relative to equilibrium (pink marks in Fig.~S5), which is consistent with the observed increase in capillary wave amplitudes under these conditions.
These trends are reversed when the chemical drive is applied in the opposite direction (teal marks in Figs.~\ref{fig:figure2.5}B and S4).

Taken together, these simulation results indicate that both spatial and compositional fluctuations of phase-separated interfaces can deviate from thermal equilibrium under conditions where nonequilibrium reactions drive steady-state mesoscopic fluxes.
Consequently, we anticipate that equilibrium approximations of phase-separated interfaces are unlikely to quantitatively describe the phase behavior of this model fluid under generic nonequilibrium conditions.

\section*{A microscopic theory predicts nonequilibrium phase coexistence}

Motivated by these observations, we propose a first-principles theory that combines a microscopic treatment of the interface with a mesoscopic description of the coexisting bulk phases (see \textit{Materials and Methods}).
This approach allows us to predict the conditions for nonequilibrium phase coexistence directly from the microscopic details of the molecular interactions and reaction rate constants, without relying on local equilibrium assumptions or equilibrium free-energy densities.
We model the dilute phase using a standard reaction--diffusion equation for an ideal mixture of B and I-state molecules at steady state,
\begin{equation}\label{eq:reaction-diffusion}
  0 = \xi^2\nabla^2\delta\rhoBI - \delta\rhoBI,
\end{equation}
where $\delta\rhoB$ and $\delta\rhoI$ represent the differences in the dilute-phase concentrations of the B and I-state molecules relative to the far-field values, $\rhoBv$ and $\rhoIv$, respectively.
The reaction--diffusion length $\xi$ and the ratio $\rhoBv/\rhoIv$ can be predicted from the mobility, $\Lambda$, and the total reaction rates, $\kIB = \kIB^{\text{p}} + \kIB^{\text{d}}$ and $\kBI = \kBI^{\text{p}} + \kBI^{\text{d}}$, in the dilute limit.
\eqref{eq:reaction-diffusion} accurately describes the concentration profile in the dilute phase (Figs.~\ref{fig:figure2}A and S5) but requires two boundary conditions at the phase-separated interface.
To this end, we describe the microscopically sharp interface in terms of a microscopic probability distribution $p(i,\nB)$ representing the probability that a tagged lattice site is occupied by a molecule or implicit solvent of species $i$ while being surrounded by $\nB$ B-state neighbors.
This distribution must be stationary at steady state, $\dot p = 0$.
For simplicity, we do not explicitly model the condensed phase, and we assume that all fluxes in the condensed phase are negligible as observed in our simulations; however, our approach could be generalized to include such effects as well under appropriate conditions (see SI).

We then connect the microscopic description of the interface to the mesoscopic description of the dilute phase by employing a perturbative Fixed Local Environment approXimation (FLEX)~\cite{cho2023nucleation,cho2023interface}, in which the configuration around a tagged lattice site at the interface is assumed to change slowly compared to single-site fluctuations.
This approximation implies that the stationary condition for $p(i,\nB)$ can be solved separately for each value of $\nB$ and that the marginal distribution, $p(\nB)$, is unperturbed by the chemical drive.
We solve for the dilute-phase concentrations adjacent to the interface, $\rhoBI(R+)$, by matching the microscopic and mesoscopic diffusive fluxes,
\begin{equation}\label{eq:matched_diffusive_flux}
  \left\langle \jB^\text{FLEX}[\pB(\nB),\pI(\nB)] \right\rangle = -(\Lambda/4) \nabla \rhoB \big|_{R+},
\end{equation}
where $p_{\text{B/I}}(\nB)$ is the conditional distribution given a local environment with $\nB$ B-state neighbors and $\langle \cdot \rangle$ indicates an average over $\nB$ weighted by $p(\nB)$.
Finally, we establish phase coexistence using a kinetic definition of the supersaturation that exploits the particle--hole symmetry of the lattice gas~\cite{cho2023nucleation},
\begin{equation}\label{eq:pB2}
  \ln \left[ \langle\pB\rangle / \left(1 - \langle\pB\rangle\right) \right] - \beta\sigma R^{-1} = 0,
\end{equation}
where the second term is the Gibbs--Thomson correction for a curved interface with nonequilibrium interfacial tension $\sigma$~\cite{porter2021phasetransformation}.
\eqref{eq:pB2} is equivalent to the equal pressure condition for phase coexistence at equilibrium and, under nonequilibrium conditions, ensures that $\langle\pB\rangle = \nicefrac{1}{2}$ at the Gibbs dividing surface in the case of a flat interface (i.e., $R \rightarrow \infty$).
\eqsref{eq:matched_diffusive_flux} and (\ref{eq:pB2}) constitute the interfacial boundary conditions needed to determine the dilute-phase concentration profile, and thus to predict the total far-field concentration $\rhov$, given the reaction rates specified by \eqsref{eq:forward_rates} and (\ref{eq:local_detailed_balance}) and the mobility $\Lambda$.
Importantly, nonzero steady-state diffusive fluxes imply that reactive fluxes take place on the interface due to mass conservation and the stationary condition for $p(i,\nB)$.

\begin{figure*}
    \centering
    \includegraphics[width=0.75\textwidth]{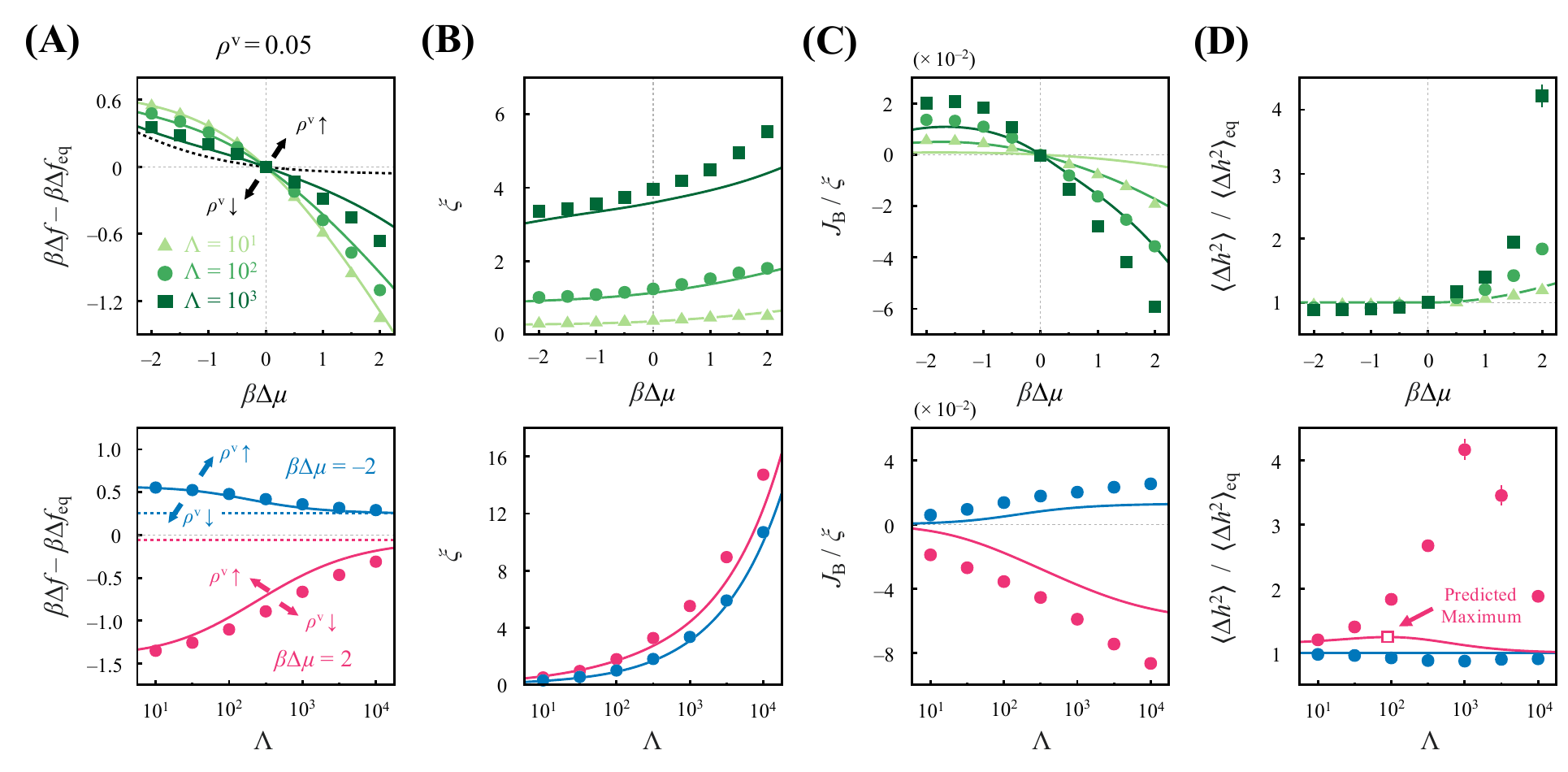}
    \caption{
    \textbf{Predictions of a microscopic--mesoscopic hybrid theory agree with simulations of nonequilibrium phase coexistence.}
    Simulation results (marks) and theoretical predictions (solid lines) at fixed $\Lambda$ (upper row) or $\beta\Dmu$ (lower row) for a macroscopically flat interface (i.e., $R \rightarrow \infty$; cf.~\figref{fig:figure1}C).
    (A) Phase diagrams showing coexistence curves given a fixed far-field total concentration in the dilute phase, $\rhov = 0.05$. Moving away from these coexistence curves either increases $(\rhov\!\uparrow)$ or decreases $(\rhov\!\downarrow)$ the far-field concentration. Dotted lines show the predictions of an alternative theory that assumes equilibrium boundary conditions at the interface (see SI).  The subscript ``eq'' indicates the value at equilibrium ($\Dmu = 0$).
    (B) The reaction--diffusion length in the dilute phase, $\xi$,
    (C) the spatially averaged net diffusive flux density near the interface, $\JB / \xi$, and
    (D) the change in the mean-squared interfacial height fluctuations, $\langle \Dh^2 \rangle$, relative to equilibrium.
    All data and predictions shown in panels B--D are made along the coexistence lines shown in panel A.
    In the upper row of panel D, only the prediction for $\Lambda=10^2$ is presented to show the qualitative trend.
    }
    \label{fig:figure3}
\end{figure*}

Comparisons between theoretical predictions and simulation results over a wide range of $\Dmu$ and $\Lambda$ values demonstrate the unique advantages of this approach.
We first examine the case of a macroscopically flat interface, which we simulate using direct-coexistence simulations in a slab geometry~\cite{cho2023interface}, in \figref{fig:figure3}.
In \figref{fig:figure3}A, we predict nonequilibrium coexistence curves in which the total far-field concentration is held constant.
These curves describe how the internal energy difference between the B and I states, $\Df$, must be tuned in order to maintain a constant far-field concentration of $\rhov = 0.05$ in the dilute phase, which guarantees that the dilute phase is always approximately ideal.
We also plot the predictions of an alternative theoretical approach in which the interfacial boundary conditions for \eqref{eq:reaction-diffusion} impose equilibrium phase coexistence by requiring constant chemical potentials and pressures across the sharp interface~\cite{wurtz2018chemical,bauermann2022energy} (see SI).
By contrast with \eqsref{eq:matched_diffusive_flux} and (\ref{eq:pB2}), these equilibrium boundary conditions do not predict $\Lambda$-dependent far-field concentrations and thus do not agree with the simulation data in general.
However, the two theories yield similar results in the limit $\Lambda\rightarrow\infty$, since the phase-separated fluid approaches a global effective equilibrium when the timescale for reactions diverges relative to the timescale for diffusion (see \textit{Materials and Methods}).
Analogous predictions in which we hold $\Df$ constant and predict $\rhov$ at phase coexistence are presented in Fig.~S6.

We also obtain good agreement between the predicted and measured reaction--diffusion length in the dilute phase, $\xi$ (\figref{fig:figure3}B), and the spatially averaged net diffusive flux density near the interface, $\JB/\xi \equiv \xi^{-1}\int_0^\infty \jB\,dz$, at phase coexistence (\figref{fig:figure3}C).
Although $\xi$ is always positive, $\JB$ is identically zero at equilibrium ($\Dmu=0$) since net diffusive fluxes and their associated concentration gradients are inherently nonequilibrium phenomena.
Moreover, the net diffusive flux reverses direction when the sign of $\Dmu$ is flipped, corresponding to a change in the direction of the reactive fluxes on the interface.
Further comparisons between theory and simulation results are shown for other concentration-dependent reaction schemes in Fig.~S7.

\section*{Nonequilibrium Interfacial Fluctuations are Coupled to Steady-state Fluxes}

Crucially, our theory also predicts nonequilibrium variations in the fluctuations at the interface.
Within the FLEX approximation, we find that the steady-state conditional distribution $\pBI(\nB)$ does not follow the Boltzmann distribution when $\jB(R+)/\Lambda \ne 0$.
This prediction, which is consistent with the simulation results shown in \figref{fig:figure2.5}B, can be understood by noting that both $p(i,\nB)$ and the dilute-phase concentration profiles are uniquely determined by our theory, so that further imposing the Boltzmann distribution at the interface leads to an overdetermined system of equations.
The only way that $p_{\text{B/I}}(\nB)$ can satisfy the Boltzmann distribution for all values of $\nB$ within the FLEX approximation is if the entire fluid is described by a single effective equilibrium model, which occurs when $\jB(R+)/\Lambda = 0$ and the dilute-phase concentration gradient vanishes (see SI for further discussion).

This prediction of our theory also suggests why the interfacial capillary fluctuations, which are related to the marginal distribution $p(\nB)$, can differ from equilibrium when $\jB(R+)/\Lambda \ne 0$, as shown in \figref{fig:figure2.5}A.
Although the FLEX approximation does not resolve the full distribution $p(i,\nB)$, we can predict this effect qualitatively by computing an effective ``single-bond'' nearest-neighbor interaction energy for an adsorbed B-state molecule ($\nB=1$) and then determining $\langle \Dh^2 \rangle$ from an equilibrium model of capillary waves~\cite{cho2023interface} (see \textit{Materials and Methods}).
Because interfacial height fluctuations decrease monotonically with the effective single-bond interaction strength~\cite{saito1996statistical}, weakening the effective interactions leads to larger $\langle \Dh^2 \rangle$ and vice versa.
Comparisons between this prediction and simulation results reveal similar qualitative dependencies on $\Dmu$ and $\Lambda$ (\figref{fig:figure3}D), whereby $\langle\Dh^2\rangle$ increases monotonically with respect to $\Dmu$ and goes through a maximum value with respect to $\Lambda$ when $\Dmu > 0$.
Specifically, $\langle \Dh^2 \rangle$ increases with $\Lambda$ in the reaction-dominant (small $\Lambda$) regime, but necessarily approaches the equilibrium value, $\langle \Dh^2 \rangle_{\text{eq}}$, in the diffusion-dominant (large $\Lambda$) regime where $\jB(R+)/\Lambda \rightarrow 0$ (see \textit{Materials and Methods}).
Nonetheless, the FLEX approximation tends to predict lower values of $\langle \Dh^2 \rangle$ than what is observed in simulations.
One cause of this underprediction is likely a breakdown in the timescale separation assumption that underlies FLEX, particularly when $\Lambda$ is large.
Coupling between fluctuations at a tagged lattice site and fluctuations in its local environment would therefore have to be taken into account to quantitatively match the peak values of $\Dh$ that are seen in simulations.
Moreover, the capillary fluctuation spectra show greater deviations from the equipartition $q^{-2}$ scaling at small $q$ when $\langle \Dh^2 \rangle \gg \langle \Dh^2 \rangle_{\text{eq}}$ (Fig.~S8), which cannot be explained by an effective equilibrium model of capillary waves and thus also contributes to the quantitative underprediction of $\langle \Dh^2 \rangle$ in \figref{fig:figure3}D.

Overall, the key insight from our theoretical approach is that mesoscopic fluxes and nonequilibrium interfacial properties both arise from non-Boltzmann statistics at phase-separated interfaces.
Whereas altered capillary waves are a direct manifestation of nonequilibrium interfacial fluctuations, mesoscopic fluxes in the dilute phase indirectly indicate the presence of microscopic fluxes and non-Boltzmann statistics at the interface.
Consequently, our theory predicts that nonequilibrium interfacial properties should be expected in this model whenever steady-state concentration gradients appear in the coexisting bulk phases.
Moreover, driving the fluid away from equilibrium in a manner that does not generate non-Boltzmann statistics, for example by choosing an $\nB$-independent form for $\kIB^\text{d}$ in \eqref{eq:forward_rates}, generates neither nonequilibrium interfacial properties~\cite{cho2023nucleation} nor mesoscopic fluxes (Fig.~S9).

\section*{Nonequilibrium Interfacial Effects Tune Droplet Sizes}

\begin{figure*}
    \centering
    \includegraphics[width=0.75\textwidth]{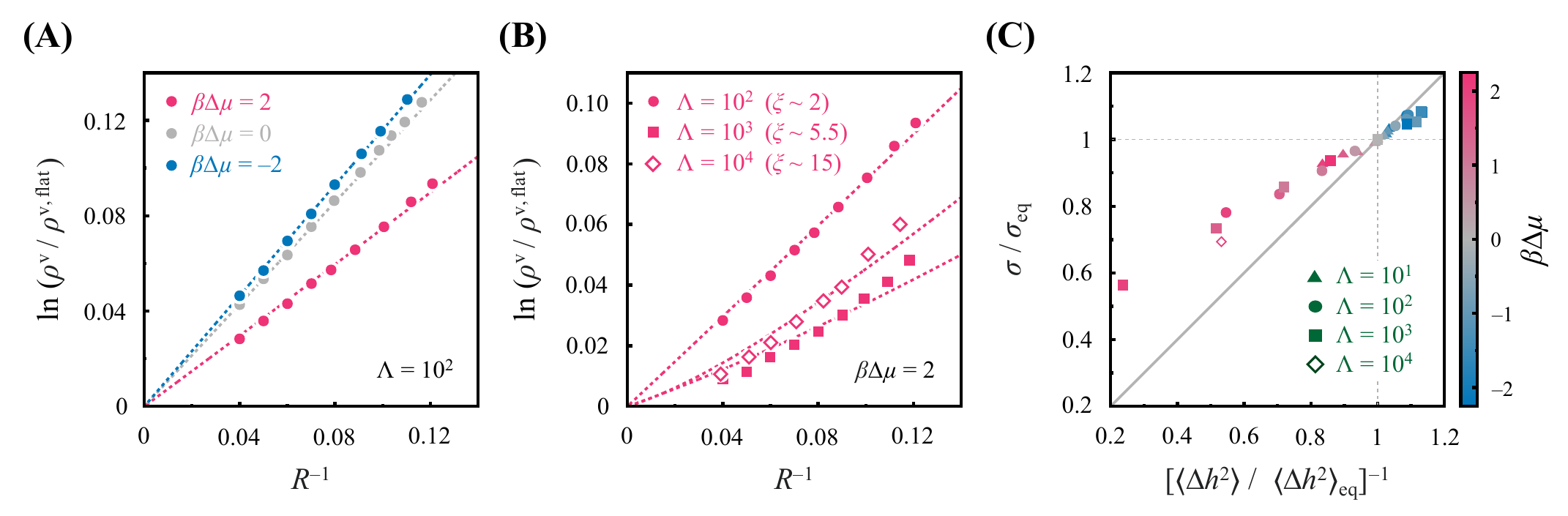}
    \caption{\textbf{The size scaling of condensed-phase droplets is governed by nonequilibrium interfacial effects.}
    (A,B)~Changes in the far-field total concentration in the dilute phase relative to a macroscopically flat interface, $\rhov / \rhovf$, measured in simulations of finite-size droplets (cf.~\figref{fig:figure1}D).
    Simulation data obtained by varying the total concentration on a lattice with fixed dimensions (marks) are compared to the theory, in which the interfacial tension, $\sigma$, is the sole fitting parameter (dotted lines).
    The corresponding flat-interface coexistence curves, where $R^{-1}\rightarrow 0$ and $\rhov\rightarrow\rhovf$, are shown in \figref{fig:figure3}A.
    (A)~Droplet size scaling in the regime $\xi \sim 2 \ll R$.
    Data are shown for the conditions examined in \figref{fig:figure2.5}: $\beta\Dmu = -2$ (teal), $0$ (gray), and $2$ (pink) at constant $\Lambda = 10^2$, with $\rhovf=0.05$.
    (B)~Droplet size scaling in the regime $\xi \gtrsim R$.
    Data are shown at constant $\beta\Dmu$ and $\rhovf=0.05$ with variable $\Lambda = 10^2$ (circles), $10^3$ (squares), and $10^4$ (diamonds), resulting in dilute-phase reaction--diffusion lengths of approximately $\xi\sim2$, $5.5$, and $15$, respectively.
    (C)~Comparison of the nonequilibrium interfacial tension inferred from droplet size scaling, $\sigma$, and the mean-squared capillary fluctuations, $\langle\Dh^2\rangle$, of a macroscopically flat interface under otherwise equivalent conditions.
    For all points shown, $\rhovf = 0.05$.
    The gray solid line indicates the equipartition prediction $\sigma/\sigma_\text{eq} = [\langle\Dh^2\rangle / \langle \Dh^2 \rangle_\text{eq}]^{-1}$, where the subscript ``eq'' indicates the value at equilibrium.
    }
    \label{fig:figure4}
\end{figure*}

Having established the conditions for nonequilibrium phase coexistence in the limit of a macroscopically flat interface, we turn our attention to finite-size condensed-phase droplets in \figref{fig:figure4}.
In these simulations, the droplet radius $R$ is controlled indirectly by changing the total concentration of molecules on a lattice with fixed dimensions, causing the droplet to grow or shrink to maintain phase coexistence at steady state.
However, due to the finite curvature at the droplet interface, the coexistence conditions change relative to the flat interface, leading to an increased far-field total concentration in the dilute phase, $\rhov > \rhovf$.
In what follows, we show that the dependence of $\rhov/\rhovf$ on the inverse droplet radius, $R^{-1}$, can be understood by examining each of the FLEX boundary conditions, \eqsref{eq:matched_diffusive_flux} and (\ref{eq:pB2}), in turn.

We first consider the scenario in which the dilute-phase reaction--diffusion length is small compared to the droplet radius, $\xi \ll R$.
In this case, the right-hand side of the flux boundary condition, \eqref{eq:matched_diffusive_flux}, is insensitive to $R$, since the droplet interface is nearly flat on length scales comparable to $\xi$ (see SI).
The $R$-dependence is thus contained entirely in \eqref{eq:pB2}, which relates the supersaturation of the condensed-phase droplet to the interfacial tension and the droplet radius.
As in equilibrium fluids, where the mechanical pressure $\Delta P$ exerted by a one-dimensional curved interface shifts the coexistence concentration of the dilute phase according the Young--Laplace equation $\beta\Delta P = \beta\sigma_{\text{eq}}/R$~\cite{porter2021phasetransformation}, we find that $\ln(\rhov / \rhovf)$ is linearly related to $R^{-1}$ when the dilute phase is approximately ideal (\figref{fig:figure4}A).
However, the constant of proportionality is not given by $\beta\sigma_{\text{eq}}$ under nonequilibrium conditions, but rather varies in accordance with the suppression or enhancement of capillary fluctuations (cf.~\figref{fig:figure2.5}A).
We interpret this effect as evidence of a nonequilibrium interfacial tension $\sigma \ne \sigma_{\text{eq}}$, as previously observed in simulations of droplet nucleation~\cite{cho2023nucleation}.
This effect, which must be accounted for to predict the steady-state size scaling of condensed-phase droplets, is absent from alternative theoretical approaches that treat the phase-separated interface using equilibrium approximations.

When the dilute-phase reaction--diffusion length is similar to or greater than the droplet radius, $\xi \gtrsim R$, the droplet size scaling can be more complicated due to changes in the diffusive flux relative to the flat-interface coexistence condition (\figref{fig:figure4}B).
Our theory captures this effect through the $R$-dependence of \eqref{eq:matched_diffusive_flux}, which leads to nonlinear variations in $\ln(\rhov / \rhovf)$ with respect to $R^{-1}$ when $\xi \gtrsim R$ (see SI).
At constant $\beta\Dmu$, this nonlinearity disappears both when $\Lambda\rightarrow0$, since $\xi \ll R$ in this limit, and when $\Lambda\rightarrow\infty$, since in this limit the system approaches a global effective equilibrium in which the concentration gradient in the dilute phase vanishes (Fig.~S10).
For example, for the conditions examined in \figref{fig:figure4}B, the nonlinearity is most significant near $\Lambda=10^4$, for which $\xi\sim15$ and there is clear evidence of curvature in the simulation data and theoretical predictions for $R^{-1} \gtrsim \xi^{-1} = 0.067$.
Nonetheless, the nonequilibrium interfacial tension must still be accounted for to obtain quantitative agreement between theory and simulation in this regime (Fig.~S10).

The nonequilibrium interfacial tensions that we infer from these simulations correlate strongly with the variations in the capillary fluctuations that we measure in simulations of macroscopically flat interfaces under otherwise equivalent conditions (\figref{fig:figure4}C).
Here we determine the value of the nonequilibrium interfacial tension by fitting our theory to the finite-$R$ simulation data using $\sigma$ as the sole fitting parameter (see \textit{Materials and Methods}).
We then compare the deviation from equilibrium, $\sigma/\sigma_{\text{eq}}$, to the deviation in the mean-squared capillary fluctuations, $\langle\Dh^2\rangle / \langle\Dh^2\rangle_{\text{eq}}$.
Importantly, we observe a consistent linear trend for all finite-$R$ droplets conducted in both the $\xi\ll R$ and $\xi \gtrsim R$ regimes, suggesting that the inferred values of $\sigma$ can indeed be interpreted in terms of nonequilibrium variations in capillary effects.
However, the slope of the linear trend in \figref{fig:figure4}C is not unity, as might be expected on the basis of the equipartition theorem.
This is most likely due to the breakdown of the equilibrium $q^{-2}$ scaling of capillary fluctuations under far-from-equilibrium conditions (Fig.~S8).
Nonetheless, these results collectively demonstrate that our theoretical approach provides a comprehensive description of droplet phase coexistence under a wide range of nonequilibrium conditions.

\section*{Discussion}

In this paper, we show that phase coexistence and mesoscopic fluxes in chemically driven fluids are governed by nonequilibrium interfacial effects.
Molecular-level simulations reveal that non-Boltzmann interfacial statistics and nonequilibrium capillary fluctuations emerge together with mesoscopic fluxes, which are confined near the interface.
Inspired by these observations, we propose a nonequilibrium theory of phase coexistence that connects a microscopic description of the interface to a mesoscopic model of the coexisting bulk phases.
This theory predicts accurate phase diagrams for both macroscopic and finite-size droplets and explains the interdependence of non-Boltzmann statistics and mesoscopic fluxes.
Our approach contrasts with alternative theories that assume equilibrium interfacial properties and are thus only valid for reactions that are weakly driven or extremely slow by comparison to molecular diffusion.
Taken together, our results highlight the importance of nonequilibrium behaviors at all length scales in fluids driven by molecular-scale reactions.

This study clarifies the extent to which various properties of chemically driven fluids can be considered to be at effective equilibria.
Far from the phase-separated interface, the bulk-phase entropy production is dominated by futile cycles, which average out when we coarse-grain to obtain mesoscopic properties~\cite{Seifert2019Inference}.
Each bulk phase may therefore be reasonably described at a mesoscopic level in terms of an effective equilibrium.
This observation also explains why the total entropy production rate provides little insight into the phase behavior and is insufficient to predict the existence of mesoscopic fluxes in general.
By contrast, effective equilibrium models do not provide accurate descriptions of the near-interfacial region, since the inhomogeneous local environments near a phase-separated interface make it impossible to define a consistent Boltzmann distribution.
Although we present results for a particular chemical reaction scheme in the main text, these central conclusions can be straightforwardly extended to alternative concentration-dependent reaction schemes with one or more driven pathways.

The theory developed in this work succeeds by combining models of inhomogeneous fluids at different length scales.
In general, mesoscopic reaction--diffusion models of chemically driven fluids~\cite{berry2018physical,weber2019physics,Julicher2024Droplet,shelest2024interface} require additional assumptions regarding the validity of equilibrium free energy functionals and the nature of phase-separated interfaces in order to provide a complete description of phase coexistence.
Although near-equilibrium assumptions are appropriate for small driving forces ($|\beta\Dmu| \ll 1$) or shallow concentration gradients ($\Lambda\rightarrow\infty$), our simulations demonstrate that the predictions of near-equilibrium theories can fail qualitatively under more general conditions.
Our theoretical approach solves this problem by treating the microscopically sharp interface as a boundary condition via the FLEX framework, which accounts for the molecular interactions and the details of the reaction scheme without assuming an effective equilibrium at any length scale.
Nonetheless, because the approximations invoked by FLEX are most effective when the capillary fluctuations are small, further work will be needed to improve the predictions of our theory near critical points.
Future studies will also be needed to pinpoint the precise relationship between the nonequilibrium interfacial tension and the capillary fluctuation spectrum under far-from-equilibrium conditions.

Finally, we emphasize that our simulation results and theoretical predictions are crucial for understanding the nucleation and coarsening behavior of nonequilibrium biomolecular condensates, as well as their ability to function as localized centers for chemical reactions~\cite{schuster2021condensate}.
Nonequilibrium interfacial properties in particular strongly affect the kinetics of nucleation by controlling the height of the nucleation barrier that must be surmounted in order for a small droplet to grow~\cite{cho2023nucleation,Zwicker2023nucleation}.
Accurately describing phase-separated interfaces is also likely to be important in the context of droplet coarsening~\cite{zwicker2015suppression,wurtz2018chemical} and droplet fusion, as these processes are driven by the minimization of the interfacial free energy in equilibrium fluids~\cite{berry2018physical,weber2019physics}.
In addition, we anticipate that nonequilibrium interfacial fluctuations will affect the adsorption of surfactant-like molecules~\cite{sanders2020competing,Li2024Surfactant}, the transport of metabolites~\cite{zhao2019metabolic,riback2023viscoelasticity}, and the catalytic properties of condensate interfaces~\cite{king2024catalytic}.
These wide-ranging implications underscore the need for quantitative experimental measurements of nonequilibrium interfacial properties, such as the measurements of droplet size scaling that we have prototyped via simulation in this work, which could be used to test our predictions for chemically driven fluids, especially in the important case of biomolecular condensates \textit{in vivo}.

\section*{Materials and Methods}
  
\textbf{Simulation Model---}
  We simulate a minimal model of a chemically driven fluid using a two-dimensional square lattice with periodic boundaries, in which the total number of molecules is conserved.
Chemical reactions and molecular diffusion are simulated via the kinetic Monte Carlo algorithm~\cite{gillespie2007stochastic} following the kinetic scheme described in the main text, with $\Dmu$ held constant over the course of each simulation.
For flat-interface simulations, we initialize the system with a $50 \times 50$ block of B-state molecules at the center of a $50 \times 200$ lattice.
For finite-droplet simulations, we initialize the system of size $L\times L$ by filling all lattice site within a distance $R$ from the center with B-state molecules, where $R\in[8.5,25]$ and $L\in[101,171]$ depending on the simulation parameters.
In both cases, additional molecules are randomly distributed in the dilute phase to initialize the simulation close to the steady-state dilute-phase concentration $\rhov$.
The system is then relaxed to steady state over $5\times10^5$ sweeps of events, after which measurements of steady-state quantities are accumulated over another $10^6$ sweeps.
Samples are collected every $10^2$ sweeps.
The reaction--diffusion length $\xi$ is measured directly in simulations of the dilute phase only (Fig.~S3).

To define the instantaneous interface in a microscopic configuration obtained via simulation, we first identify the condensed phase as the largest contiguous cluster of B-state molecule-occupied lattice sites and the dilute phase as the largest contiguous cluster of either implicit solvent-occupied or I-state molecule-occupied lattice sites.
We then define the instantaneous interface as the set of all lattice sites in each phase that neighbor lattice sites in the other phase.
Consequently, lattice sites at the interface can have $\nB \in \{1,2,3\}$.
We obtain the time-averaged mesoscopic concentration profiles $\rhoBI$ from simulations by averaging the position of the instantaneous interface with respect to either $z$ or $r$ at steady state, where we first compute the center of mass of the condensed-phase cluster in each instantaneous configuration to define the origin, $r = 0$.
The Gibbs dividing surface is then defined from the concentration profile where $\rhoB = \nicefrac{1}{2}$.
In flat-interface simulations, interfacial height fluctuations, $\Delta h$, are defined as the distance from the instantaneous interface to the Gibbs dividing surface.
We compute $\langle \Dh^2 \rangle$ by averaging over the entire interface in each of the microscopic configurations sampled at steady state.
The power spectrum of height fluctuations, $\langle |\Dh(q)|^2 \rangle$, is computed by first calculating $\Dh(q)$ for each microscopic configuration via a discrete Fourier transform and then taking the average over the spectra sampled at steady state.

The total entropy production rate density, $\dot{\Sigma}$, is calculated from a trajectory of duration $T$, comprising a total of $N$ reactive or diffusive events that occur at a particular lattice site $i$,
\begin{equation}
  \textstyle
    \dot{\Sigma}(i) \equiv T^{-1} k_\text{B}\sum_{n=1}^{N}  \ln (k_n / \tilde k_n),
\end{equation}
where $k_\text{B}$ is Boltzmann constant and $k_n$ and $\tilde k_n$ are the actual and time-reversed rates, respectively, associated with event $n$ at lattice site $i$~\cite{van2015ensemble}.
The mesoscopic diffusive flux density is measured by averaging the net number of diffusive events moving right, $N^\rightarrow_{\text{B/I}}$, versus left, $N^\leftarrow_{\text{B/I}}$, at lattice site $i$ within the simulation duration $T$,
\begin{equation} \label{eq:jBI}
    \jBI(i) = \left[N^\rightarrow_{\text{B/I}}(i) - N^\leftarrow_{\text{B/I}}(i)\right] / T.
\end{equation}
The mesoscopic reactive flux density in the $\BI$ direction, ${\sBI - \sIB}$, at lattice site $i$ is similarly defined as
\begin{equation} \label{eq:sBI}
    \sBI(i) - \sIB(i) = \left[N_\BI(i) - N_\IB(i)\right] / T,
\end{equation}
where $N_\BI$ and $N_\IB$ are the numbers of $\BI$ and $\IB$ chemical reactions, respectively, occurring at lattice site $i$ within the simulation duration $T$.
The mesoscopic diffusive and reactive fluxes in flat-interface simulations are averaged perpendicular to the $z$ axis after aligning the microscopic configurations with respect to the center of mass of the condensed phase.
In the curved-interface simulations, the concentration profiles and flux densities are measured along the principal axes passing through the center of mass of the droplet in accordance with the four-fold symmetry of the square lattice.
When decomposing the mesoscopic fluxes into contributions from the bulk phases and interface in \figref{fig:figure2}B,C, we attribute the contribution to the total fluxes in \eqsref{eq:jBI} and (\ref{eq:sBI}) to the interface if the lattice site $i$ is part of the instantaneous interface, as defined above, and otherwise to one of the bulk phases.
The dilute-phase contribution to the B-state concentration profile in \figref{fig:figure2}A is also identified by the same method.
All simulation data presented in the main text are averaged over 96 independent simulations.
Uncertainties associated with simulation measurements are comparable to or smaller than the size of symbols in all figures unless indicated explicitly.
Simulation codes are available at \texttt{\small https://github.com/wmjac/chem-driven-lattice-fluid}.

\vskip1ex
\noindent
\textbf{Microscopic--Mesoscopic Theory of Nonequilibrium Phase Separation---}
Here we summarize our theoretical approach, which combines a microscopic treatment of the interface with a mesoscopic description of the dilute phase.
For a detailed derivation and discussion, we refer the reader to the SI.
The alternative sharp-interface theory that assumes equilibrium boundary conditions at the interface~\cite{wurtz2018chemical,bauermann2022energy}, whose predictions are shown by dotted lines in \figref{fig:figure3}A, is also described in the SI.

The dilute phase is governed by \eqref{eq:reaction-diffusion} for $r > R$, which predicts that $|\delta\rhoBI|$ is greatest where the dilute phase contacts the interface at $r = R+$ and decays to zero according to the length scale $\xi$.
Assuming that the dilute phase is ideal (so that $\nB = 0$), we predict
\begin{equation}
    \xi = \sqrt{D/[\kBI(\nB=0)+\kIB(\nB=0)]},
\end{equation}
where $D = \Lambda/4$ is the self-diffusion coefficient on the two-dimensional square lattice in the ideal limit and $\kBI$ and $\kIB$ represent the summed rate constants over the passive and driven pathways.
Solving \eqref{eq:reaction-diffusion} yields the diffusive flux densities,
\begin{align}\label{eq:diffusive_flux_density_dilute}
  \jBI(r) &= -D\nabla\rhoBI \nonumber \\
  &= \dfrac{D[\rhoBI(R+)-\rhoBI^\text{v}]}{\xi}\dfrac{K_1(\xi^{-1}r)}{K_0(\xi^{-1}R)},
\end{align}
where $K_n$ is the $n$th order modified Bessel function of the second kind.
\eqref{eq:diffusive_flux_density_dilute} applies to the flat interface in the macroscopic limit, $R\rightarrow\infty$, but with both $K_n$'s replaced with exponential functions.

We next describe the microscopic interface by employing a Fixed Local Environment approXimation (FLEX)~\cite{cho2023nucleation,cho2023interface}, which assumes that the marginal distribution $p(\nB)$ is unperturbed by nonequilibrium driving.
This approach allows us to compute the steady-state probabilities $\pB(\nB)$ and $\pI(\nB)$ of finding B and I-state molecules, respectively, at a tagged lattice site on the interface with $\nB$ B-state nearest neighbors by accounting for chemical reactions and diffusive exchange with the dilute phase.
At steady state,
\begin{align}\label{eq:balance_FLEX}
  \dot{p}_\text{B}(\nB) = 0 = [&\pI(\nB)\kIB(\nB) \nonumber \\
    &- \pB(\nB)\kBI(\nB)] - \jB^\text{FLEX}(\nB),
\end{align}
where the term in brackets is the net reactive flux density, ${\sIB(\nB)-\sBI(\nB)}$, and the net diffusive flux density is
\begin{equation}\label{eq:diffusive_flux_FLEX}
  \begin{aligned}
    \jB^\text{FLEX}(\nB) = D\{&\pB(\nB) e^{\nB\beta\epsilon}[1-\rhoB(R+)-\rhoI(R+)] \\ &- \rhoB(R+)[1-\pB(\nB)-\pI(\nB)]\}.
  \end{aligned}
\end{equation}
\eqsref{eq:balance_FLEX} and (\ref{eq:diffusive_flux_FLEX}), along with analogous equations for $\dot \pI$ and $\jI^{\text{FLEX}}$, uniquely determine the conditional distribution $p_{\text{B/I}}(\nB)$ for each value of $\nB$ on the interface.
The interfacial boundary conditions \eqsref{eq:matched_diffusive_flux} and (\ref{eq:pB2}), along with conservation of mass at steady state, $\jB + \jI = 0$, and the far-field condition for an ideal dilute phase, $\rhoB^{\text{v}} / \rhoI^{\text{v}} = \kIB(\nB=0) / \kBI(\nB=0)$, then uniquely determine the four unknowns $\rhoBI(R+)$ and $\rhoBI^{\text{v}}$ at nonequilibrium phase coexistence.

Within the FLEX approximation, we predict qualitative variations in the interfacial height fluctuations by computing an effective nearest-neighbor interaction energy, $\tilde{\epsilon}$, for an adsorbed B-state molecule,
\begin{equation}
  \beta\tilde\epsilon = \ln\left[\pB(\nB=1)/(1-\pB(\nB=1))\right],
  \label{eq:effective-epsilon}
\end{equation}
and then applying an equilibrium theory of capillary fluctuations~\cite{saito1996statistical} in which $\langle \Dh^2 \rangle \propto \sinh^2(\beta\tilde\epsilon/4)$ at phase coexistence.
Empirical values of $\sigma$ are extracted from simulations of finite-size droplets carried out with a constant simulation volume but varying overall concentration, resulting in a range of steady-state droplet radii (\figref{fig:figure4}A,B).
The far-field total concentration $\rhov$ for each simulation is determined by averaging the concentration far from the droplet interface, i.e., $r > R+2\xi$.
Finally, we determine $\sigma$ by first predicting the conditions for phase coexistence in the limit $R\rightarrow\infty$ and then fitting the theory to the simulation data $\ln(\rhov/\rhovf)$ versus $R^{-1}$ using $\sigma$ as the sole fitting parameter (\figref{fig:figure4}C).

The dilute-phase concentration gradient is predicted by this theory to vanish in both the reaction-dominant limit ($\Lambda\rightarrow 0$) and in the diffusion-dominant limit ($\Lambda\rightarrow\infty$) for finite $\beta\Dmu$.
In the diffusion-dominant limit, $\JB/\xi$ is nonzero (\figref{fig:figure3}C), but $\jB(r=R+)/D \rightarrow 0$ since $\jB(r=R+)$ must be finite to satisfy the FLEX boundary condition given by \eqref{eq:matched_diffusive_flux}.
Consequently, this limit corresponds to an effective equilibrium and the interfacial statistics approach the Boltzmann distribution, implying $\langle\Dh^2\rangle \rightarrow \langle\Dh^2\rangle_{\text{eq}}$ (\figref{fig:figure3}D) and $\sigma \rightarrow \sigma_{\text{eq}}$.
In the reaction-dominant limit, both $\jB(r=R+) \rightarrow 0$ and $\JB/\xi \rightarrow 0$ (\figref{fig:figure4}C).
However, $\jB(r=R+) = 0$ implies that $\sIB(\nB)=\sBI(\nB)$ for every value of $\nB$ on the interface according to \eqref{eq:balance_FLEX}, leading to non-Boltzmann statistics and thus $\langle\Dh^2\rangle\ne\langle\Dh^2\rangle_{\text{eq}}$ (\figref{fig:figure3}D) and $\sigma \ne \sigma_{\text{eq}}$ for the reaction scheme presented in \eqref{eq:forward_rates}.
A detailed discussion of these limiting behaviors is provided in the SI.\\~

\begin{acknowledgments}
This work is supported by the National Science Foundation (DMR-2143670).
\end{acknowledgments}

\providecommand{\noopsort}[1]{}\providecommand{\singleletter}[1]{#1}%

\end{document}

\title{Supplementary Information for ``Interfacial Effects Determine Nonequilibrium Phase Behaviors in Chemically Driven Fluids''}

\author{Yongick Cho}
\affiliation{Department of Chemistry, Princeton University, Princeton, NJ 08544, USA}
\author{William M. Jacobs}
\email{wjacobs@princeton.edu}
\affiliation{Department of Chemistry, Princeton University, Princeton, NJ 08544, USA}

\maketitle

\onecolumngrid

In this supporting material, we provide a complete description of the microscopic--mesoscopic theory of phase coexistence in chemically driven fluids that is introduced in the main text, and we compare this theory with an alternative approach in which the phase-separated interface is assumed to be at thermal equilibrium.

\subsection*{Mesoscopic Description of the Dilute Phase}

We consider a two-dimensional system at a steady state in which a condensed-phase droplet with radius $R$ is centered at the origin, $r = 0$.
The coordinate $r$ thus indicates the distance from the center of mass of the droplet.
At a mesoscopic level, the dilute phase in the region $r > R$ can be described by coupled reaction--diffusion equations at steady state,
\begin{subequations}
  \label{eq:eq1}
\begin{align}
    &0 = \dot{\rhoB} = \DB\nabla^2\rhoB - \sBI(\rhoB,\rhoI) + \sIB(\rhoB,\rhoI)\\
    &0 = \dot{\rhoI} = \DI\nabla^2\rhoI + \sBI(\rhoB,\rhoI) - \sIB(\rhoB,\rhoI),
\end{align}
\end{subequations}
where $\rhoBI$ and $D_\text{B/I}$ are the concentration profiles and self-diffusion coefficients for the B-state and I-state molecules, respectively, and $\sBI$ and $\sIB$ are the reactive flux densities for the chemical reactions in the indicated directions.
Here we have assumed that the dilute phase is ideal, since the molecules are sparsely distributed, and that the self-diffusion coefficients are independent of concentration.
For small changes in the concentration profiles, $\delta\rhoBI \equiv \rhoBI - \rho^\text{v}_\text{B/I}$, relative to the far-field concentrations, $\rhoBI^\text{v}$, the change in reactive flux densities can be linearized with respect to $\delta\rhoBI$,
\begin{subequations}
  \label{eq:eq2}
\begin{align}
    &0 = \delta\dot{\rho}_\text{B} = \DB\nabla^2\delta\rhoB - \kTBI\delta\rhoB + \kTIB\delta\rhoI\\
    &0 = \delta\dot{\rho}_\text{I} = \DI\nabla^2\delta\rhoI + \kTBI\delta\rhoB - \kTIB\delta\rhoI,
\end{align}
\end{subequations}
where $\kTBI$ and $\kTIB$ are the linearized reaction rate constants evaluated at $\delta\rhoBI = 0$.
Summing the two equations in \eqref{eq:eq2} leads to
\begin{equation}\label{eq:Laplace}
    0 = \delta\dot{\rho}_\text{B} + \delta\dot{\rho}_\text{I} = \nabla^2(\DB\delta\rhoB + \DI\delta\rhoI).
\end{equation}
Because the system is azimuthal-symmetric and the far-field concentrations should not diverge, \eqref{eq:Laplace} implies
\begin{equation}\label{eq:eq4}
    0 = \DB\delta\rhoB + \DI\delta\rhoI.
\end{equation}
Finally, substituting $\delta\rhoI = -(\DB/\DI)\delta\rhoB$ into \eqref{eq:eq2} yields the linearized reaction--diffusion equations
\begin{equation}\label{eq:linearized_RD}
    0 = \xi^2\nabla^2\delta\rho_\text{B/I} - \delta\rho_\text{B/I},
\end{equation}
which is Eq.~(3) in the main text.
In \eqref{eq:linearized_RD}, the dilute-phase reaction--diffusion length $\xi$ is defined to be
\begin{equation}\label{eq:xi}
    \xi\equiv\sqrt{[\kTBI/\DB + \kTIB/\DI]^{-1}}.
\end{equation}

Solving \eqref{eq:linearized_RD} in the region $r > R$ gives the concentration profiles outside the droplet,
\begin{equation}\label{eq:rhoB_2D}
    \rhoBI(r) = \left[\rhoBI^+-\rhoBI^\text{v}\right]\dfrac{K_0(\xi^{-1}r)}{K_0(\xi^{-1}R)} + \rhoBI^{\text{v}},
\end{equation}
where $\rhoBI^+ \equiv \rhoBI(R+)$ is the concentration just above the droplet interface and $K_\text{n}$ is the nth order modified Bessel function of the second kind.
When the concentration profiles in \eqref{eq:rhoB_2D} are nonuniform, net molecular diffusion into or out of the droplet results in the flux densities
\begin{equation}\label{eq:diffusive_flux_density}
    \jBI(r) = -D_\text{B/I}\nabla\rhoBI = \dfrac{D_\text{B/I}[\rhoBI^+-\rhoBI^\text{v}]}{\xi}\dfrac{K_1(\xi^{-1}r)}{K_0(\xi^{-1}R)}.
\end{equation}
Eqs.~(\ref{eq:rhoB_2D}) and (\ref{eq:diffusive_flux_density}) indicate that nonequilibrium mesoscopic effects in the dilute phase are concentrated within a distance comparable to $\xi$ from the phase-separated interface.
In the limit of a macroscopically flat interface, i.e., $R\rightarrow\infty$, the curvature of the droplet interface vanishes, and Eqs.~(\ref{eq:rhoB_2D}) and (\ref{eq:diffusive_flux_density}) asymptotically approach one-dimensional exponential profiles, as expected for flat interfaces. 
From the asymptotic relation $K_\text{n}(\xi^{-1}r)\rightarrow\sqrt{\xi\pi/2r}\exp(\xi^{-1}r)$~\cite{silverman1972special}, Eqs.~(\ref{eq:rhoB_2D}) and (\ref{eq:diffusive_flux_density}) become
\begin{equation}\label{eq:rhoB_1D}
    \rhoBI(z)= \left[\rhoBI^+ - \rhoBI^\text{v}\right]\exp(-\xi^{-1}z) + \rhoBI^\text{v}
\end{equation}
and
\begin{equation}\label{eq:jB_1D}
    \jBI(z) = \dfrac{D_\text{B/I}[\rhoBI^+-\rhoBI^\text{v}]}{\xi}\exp(-\xi^{-1}z),
\end{equation}
respectively, where $z = r-R > 0$ is the distance from the interface in a one-dimensional system.
The dependence on $\sqrt{r^{-1}}$ is eliminated in this limit since $R \gg \xi$, so that $\sqrt{R/r}\approx1$ within the domain $r\in[R,R+\xi]$ in which the nonequilibrium mesoscopic effects are concentrated.

The assumption of ideal behavior in the dilute phase allows for further simplifications with regard to the self-diffusion coefficients and linearized reaction rates.
By assuming that interactions among molecules are negligible, we can approximate the number of nearest-neighbor B-state molecules as $\nB \approx 0$.
The ideality assumption therefore fixes the rate constants in the dilute phase to be $\kTBI = \kBI(\nB=0)$ and $\kTIB = \kIB(\nB=0)$, where $\kIB$ and $\kIB$ indicate the sum of the reaction rate constants over the passive and driven pathways.
Assuming negligible interactions among molecules also implies that the B and I-state molecules share a common self-diffusion coefficient, $D$, which is equal to
\begin{equation}\label{eq:goody_eq}
    \DB = \DI = D = \Lambda / 4
\end{equation}
on the two-dimensional square lattice with molecular mobility $\Lambda$.
When combined with \eqref{eq:eq4}, \eqref{eq:goody_eq} implies
\begin{equation}\label{eq:local_mass_conservation_dilute_phase}
    0 = \delta\rhoB + \delta\rhoI,
\end{equation}
resulting in a uniform total concentration profile, $\rhoB + \rhoI$, in the dilute phase.
Finally, by combining these results, we can predict the reaction--diffusion length $\xi$ from the microscopic details of the model,
\begin{equation}\label{eq:xid}
    \xi = \sqrt{\Lambda/4[\kBI(\nB=0)+\kIB(\nB=0)]}.
\end{equation}

\subsection*{Test of the Mesoscopic Theory Assumptions in the Dilute Phase}

To test the assumption of ideality in the dilute phase, we directly measure the reaction--diffusion length $\xi$ and compare with the predictions of the linearized reaction--diffusion equation.
First, we calculate the mean-squared displacement, $\langle \Delta r^2_\text{B/I} \rangle$, over a long simulation with the composition fixed to the far-field total concentrations $\rhoBv$ and $\rhoIv$.
In this simulation, chemical reactions are disallowed in order to obtain continuous trajectories of the molecules without state changes.
For the example conditions studied in Fig.~2A in the main text, we observe that the mean-squared displacement increases linearly with respect to time $t$, as expected for molecular diffusion, and we fit the self-diffusion coefficients using the relation
\begin{equation}
    D_\text{B/I} = \langle \Delta r^2_\text{B/I}(t) \rangle / 4t.
\end{equation}
However, we find that $\DB$ and $\DI$ are not identical in the simulations, as assumed in \eqref{eq:goody_eq}, since both excluded volume interactions and attractive interactions among B-state molecules reduce the diffusion coefficients (\figref{fig:SI_RD_test}A).

We then turn on chemical reactions so that the system composition, given by $\rhoB$ and $\rhoI = \rhov - \rhoB$, changes over time at the fixed far-field total concentration $\rhov$.
The reactive flux densities in both directions are determined from
\begin{subequations}
\begin{align}
    \sBI(\rhoB) &\equiv [N_\BI(\rhoB) / T(\rhoB)]/L^2\\
    \sIB(\rhoI) &\equiv [N_\IB(\rhoI) / T(\rhoI)]/L^2,
\end{align}
\end{subequations}
where $N_\BI$ and $N_\IB$ are the number of chemical reactions in each direction, $T$ is the time spent at a given composition, and $L$ is the linear dimension of the system.
The reactive flux densities change linearly with respect to small variations in the molecular concentrations around the average steady-state values, $\sBI(\rhoBv)$ and $\sIB(\rhoIv)$, from which empirical linearized reaction rate constants $\kTBI$ and $\kTIB$ are extracted.
For the conditions studied in Fig.~2A in the main text, the extracted rate constants $\kTIB = 0.983$ and $\kTBI = 2.75$ are comparable to the predicted values $\kIB(\nB=0) = 1.024$ and $\kBI(\nB=0) = 6.662$ (\figref{fig:SI_RD_test}B).
Finally, we evaluate $\xi$ from \eqref{eq:xi} using the measured values of the self-diffusion coefficients and linearized reaction rate constants.
This approach is used to determine the measured values of $\xi$ that are presented in Fig.~4B in the main text.

Comparisons of the observed and predicted concentration profiles using the empirically determined reaction--diffusion length scale confirm the applicability of the linearized reaction--diffusion equation, \eqref{eq:linearized_RD}, within the dilute phase under the conditions explored in this study.
For the purpose of this comparison, we account for the fluctuating interface, which obscures the dilute-phase concentration gradient near the interface in Fig.~2A in the main text, by measuring the concentration profiles with respect to the instantaneous interface.
Specifically, we define the coordinate $\tilde z$ to represent the distance from the instantaneous interface, measured normal to the Gibbs dividing surface, using the definition of the instantaneous interface given in the \textit{Materials and Methods} in the main text.
We then use this coordinate to compute $\delta\rhoBI(\tilde z) \equiv \rhoBI(\tilde z) - \rhov_{\text{B/I}}$ for flat-interface simulations (\figref{fig:SI_RD_test}C).
We similarly define the instantaneous position of the interface to be $\tilde R$ in curved-interface simulations, and we measure the concentration profiles with respect to $r - \tilde R$ in this case (\figref{fig:SI_RD_test}D).
In \figref{fig:SI_RD_test}C,D, the concentration profiles $\delta\rhoB(\tilde z)$ and $\delta\rhoB(r - \tilde R)$ are fitted to \eqsref{eq:rhoB_1D} and (\ref{eq:rhoB_2D}), respectively, using $\rhoB^+$ as the sole fitting parameter.
This comparison shows that the prediction of the reaction--diffusion equation captures the concentration gradient of B-state molecules near the interface using the empirically determined reaction--diffusion length $\xi$.
Moreover, this comparison shows that \eqref{eq:local_mass_conservation_dilute_phase} is reasonably accurate in the dilute phase, since the fit to the B-state concentration profile approximately predicts the I-state concentration profile except right near the interface.
Thus, the simplifications introduced by assuming the ideal limit in \eqsref{eq:linearized_RD} and (\ref{eq:local_mass_conservation_dilute_phase}) do not qualitatively affect the accuracy of the mesoscopic description of the dilute phase.

We also compare this method of computing the reaction--diffusion length to a direct measurement of the dilute-phase correlation length, which we define as the average displacement of a molecule before it undergoes a chemical reaction at steady state.
This correlation length is determined from a simulation of the dilute phase by computing
\begin{equation}\label{eq:correlation_length} \textstyle
    l \equiv \left[\sum_{n=1}^{N_\BI} l_\text{B}(n) + \sum_{n=1}^{N_\IB} l_\text{I}(n)\right] / \left(N_\BI + N_\IB\right),
\end{equation}
where $l_\text{B}$ and $l_\text{I}$ are the observed displacements of B and I-state molecules between successive $\BI$ and $\IB$ chemical reactions, respectively.
We observe a strong correlation between $\xi$ and $l$ in \figref{fig:SI_xi_direct}, which is expected since both quantities describe the same mesoscopic length scale via slightly different definitions.
In the rest of this work, however, we exclusively consider the reaction--diffusion length $\xi$ as defined in \eqref{eq:xi}, since we find that this calculation accurately describes the concentration of B-state molecules at phase-separated steady states, as shown in \figref{fig:SI_RD_test}C,D.

Lastly, we note that the simulations confirm the tight coupling between mesoscopic reactive and diffusive fluxes that are implied by the conservation equation, \eqref{eq:local_mass_conservation_dilute_phase}.
In \figref{fig:SI_coupling}, the contributions of the interface and the dilute phase to the reactive and diffusive flux densities obtained from simulations are rescaled relative to their own peak values for comparison.
Even though the two flux densities are measured separately, they share the location of their peak values and exhibit the same spatial variations, which are maximized near the interface for both flat and curved interfaces.

\subsection*{Mesoscopic Description of the Condensed Phase}

An analogous reaction--diffusion description can also be applied to the condensed phase.
In our microscopic model, molecules can only be displaced when they are adjacent to an unoccupied lattice site (i.e., a ``hole'') so that the mobility is effectively scaled by the concentration of holes, $\rhoSl$, relative to the dilute phase.
Assuming that the holes are sparsely distributed in the condensed phase, so that molecules are adjacent to at most one hole at a time, the condensed-phase self-diffusion coefficients for the B and I-state molecules are approximately
\begin{subequations}
\begin{align}
    &\DBl = (\Lambda/4)\rhoSl \zeta^{-1}\\
    &\DIl = (\Lambda/4)\rhoSl,
\end{align}
\end{subequations}
where the factor $\zeta^{-1} \equiv e^{3\beta\epsilon}$ accounts for the fact that B-state molecules that are adjacent to a hole in the condensed phase have $\nB\approx3$ B-state nearest neighbors in this approximation.
Linearizing the reaction--diffusion equations around the far-field condensed phase densities $\rhoBI^{\text{l}}$ again leads to \eqref{eq:linearized_RD}, but with $\xi$ replaced by $\xil$ in the condensed phase.
Boundary conditions for \eqref{eq:linearized_RD} are similarly required where the condensed phase contacts the interface, $\rhoBI^- \equiv \rhoBI(R-)$.

Nonetheless, in the case of the strongly phase-separated conditions that we study in this work, we observe negligible mesoscopic fluxes in the condensed phase.
By comparing $\xil$ with $\xi$,
\begin{equation}\label{eq:xi_ratio}
    \left(\dfrac{\xi^\text{l}}{\xi}\right)^2 = \rhoSl\dfrac{\kBI(\nB=0)+\kIB(\nB=0)}{\zeta\kBI(\nB=4)+\kIB(\nB=4)},
\end{equation}
where we have assumed $\nB\approx4$ in the condensed phase, we find that $(\xil)^2/\xi^2\ll1$, which implies that the range of mesoscopic fluxes in the condensed phase is much smaller than in the dilute phase.
This prediction is supported by the lack of any noticeable contribution from the condensed phase to the mesoscopic flux densities in Fig.~2B,C in the main text.
Evaluating \eqref{eq:xi_ratio} at the conditions corresponding to that figure, $(\beta\epsilon,\beta\Dmu,\Lambda,\beta\Df) = (-2.95,2,10^2,1.7337)$, in which the observed hole concentration is $\rhoSl = 2.5\times10^{-3} \approx \exp(2\beta\epsilon)$, leads to $(\xil)^2/\xi^2 = 4.2\times10^{-3}$.
We therefore ignore contributions from the condensed phase in our theory and focus only on the contributions from the dilute phase and the interface in the main text.

\subsection*{Modeling Nonequilibrium Phase Coexistence with a Sharp Interface}

We now develop a sharp-interface theory of nonequilibrium phase coexistence.
We first consider the case of a macroscopically flat interface (i.e., $R\rightarrow\infty$) and return to the case of a curved droplet interface later.
In the case of a macroscopically flat interface, the mesoscopic concentration profiles for the two coexisting phases meet at the sharp interface at $z=0$, as shown schematically in the following diagram:
\begin{center}
  \vskip1ex
  \includegraphics[width=7cm]{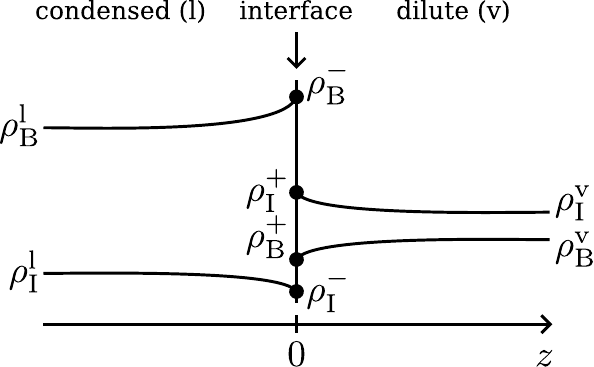}
  \vskip1ex
\end{center}
By introducing ``(non)equilibrium constants'' for the far-field boundary conditions,
\begin{subequations}
\begin{align}
  \frac{\rhoBv}{\rhoIv} = \frac{\kIB^{\circ,\text{v}}}{\kBI^{\circ,\text{v}}} \equiv \Kv \\
  \frac{\rhoBl}{\rhoIl} = \frac{\kIB^{\circ,\text{l}}}{\kBI^{\circ,\text{l}}} \equiv \Kl,
\end{align}
\end{subequations}
we can write the steady-state concentration fields in the dilute and liquid phases as
\begin{subequations}
  \label{eq:conc-fields}
\begin{align}
  \rhoB(z) &= \begin{cases} \displaystyle
    \frac{\Kl}{\zeta + \Kl} (\rhoB^- + \zeta \rhoI^-) + \left[ \frac{\zeta}{\zeta + \Kl} \rhoB^- - \frac{\zeta \Kl}{\zeta + \Kl} \rhoI^- \right] e^{z / \xil} & z < 0 \\ \displaystyle
    \frac{\Kv}{1 + \Kv} (\rhoB^+ + \rhoI^+) + \left[ \frac{1}{1 + \Kv} \rhoB^+ - \frac{\Kv}{1 + \Kv} \rhoI^+ \right] e^{-z / \xi} \quad & z > 0
  \end{cases} \\
  \rhoI(z) &= \begin{cases} \displaystyle
    \frac{1}{\zeta + \Kl} (\rhoB^- + \zeta \rhoI^-) - \dfrac{1}{\zeta}\left[  \frac{\zeta}{\zeta + \Kl} \rhoB^- -\frac{\zeta \Kl}{\zeta + \Kl} \rhoI^-\right] e^{z / \xil} & z < 0 \\ \displaystyle
    \frac{1}{1 + \Kv} (\rhoB^+ + \rhoI^+) - \left[ \frac{1}{1 + \Kv} \rhoB^+  -\frac{\Kv}{1 + \Kv} \rhoI^+  \right] e^{-z / \xi} & z > 0.
  \end{cases}
\end{align}
\end{subequations}
From these expressions, we can determine the steady-state concentration profiles, and thus the far-field concentrations, from the boundary conditions at the interface,
\begin{subequations}
  \label{eq:coex}
\begin{align}
  \rhoBv = \frac{\Kv}{1 + \Kv} (\rhoB^+ + \rhoI^+) & \qquad
  \rhoIv = \frac{1}{1 + \Kv} (\rhoB^+ + \rhoI^+) \\
  \rhoBl = \frac{\Kl}{\zeta + \Kl} (\rhoB^- + \zeta \rhoI^-) & \qquad
  \rhoIl = \frac{1}{\zeta + \Kl} (\rhoB^- + \zeta \rhoI^-).
\end{align}
\end{subequations}
We therefore need four boundary conditions at the sharp interface to determine $\rhoBI^\pm$ at phase coexistence.

\subsection*{Nonequilibrium Interfacial Boundary Conditions Using the Fixed Local Environment ApproXimation (FLEX)}

To describe the nonequilibrium steady-state properties of the interface at a microscopic level, we utilize the Fixed Local Environment approXimation (FLEX) introduced in Refs.~\cite{cho2023nucleation,cho2023interface}.
Within the FLEX framework, we focus on a tagged lattice site that is assumed to be at steady state with respect to a frozen local environment.
Specifically, we assume that the number of nearest-neighbor B-state molecules, $\nB$, remains constant for each tagged lattice site at the interface.
This approximation is valid when the chemical reactions and diffusion at the tagged site occur much faster than the collective rearrangement of the interface.
In light of this, we expect the FLEX predictions to be most accurate in the reaction-dominant limit, $\Lambda\rightarrow0$, where the rearrangement of the interface via diffusion is suppressed.
Importantly, this approach allows reactions to take place on the interface itself, which is a qualitative distinction from the equilibrium interface approach that we discuss later.
Moreover, this approach makes it possible to resolve fluctuations with respect to the conditional distribution $\pBI(\nB)$ on the interface; however, the FLEX approximation implies that the marginal distribution $p(\nB)$ is unperturbed relative to equilibrium.

Within the FLEX approximation, we treat each local environment separately, resulting in two steady-state equations for each value of $\nB \in \{1,2,3\}$ on the interface,
\begin{subequations}
  \label{eq:flex-ss}
  \begin{align}
    \dot \pB(\nB) &= [\sIB(\nB) - \sBI(\nB)] - \jB^+(\nB) + \jB^-(\nB) = 0 \\
    \dot \pI(\nB) &= [\sBI(\nB) - \sIB(\nB)] - \jI^+(\nB) + \jI^-(\nB) = 0.
  \end{align}
\end{subequations}
Transport of molecules to the interface occurs via net diffusive fluxes, which couple the interfacial region to the two bulk phases.
Based on the microscopic dynamical scheme presented in the main text, the net fluxes in each direction involving a lattice site with $\nB$ neighbors are
\begin{subequations}
  \label{eq:FLEX-fluxes}
  \begin{align}
    \jB^+(\nB) &= D \left[ e^{\beta\epsilon \nB} \pB(\nB) (1 - \rhoB^+ - \rhoI^+) - \rhoB^+ (1 - \pB(\nB) - \pI(\nB)) \right] \\
    \jI^+(\nB) &= D \left[ \pI(\nB) (1 - \rhoB^+ - \rhoI^+) - \rhoI^+ (1 - \pB(\nB) - \pI(\nB)) \right] \\
    \jB^-(\nB) &= -D \left[ e^{\beta\epsilon \nB} \pB(\nB) (1 - \rhoB^- - \rhoI^-) - e^{3\beta\epsilon} \rhoB^- (1 - \pB(\nB) - \pI(\nB)) \right] \\
    \jI^-(\nB) &= -D \left[ \pI(\nB) (1 - \rhoB^- - \rhoI^-) - \rhoI^- (1 - \pB(\nB) - \pI(\nB)) \right],
  \end{align}
\end{subequations}
where $D = \Lambda / 4$ and $\pBI(\nB)$ indicates the conditional probability that the lattice site is in state $i$ given that it has $\nB$ neighbors.
The effect of excluded volume interactions on the diffusive flux densities is included via a mean-field approximation.
Because the reactive flux densities
\begin{subequations}
  \label{eq:rxn-rates}
  \begin{align}
    \sBI(\nB) &= \pB(\nB) \kBI(\nB) \\
    \sIB(\nB) &= \pI(\nB) \kIB(\nB)
  \end{align}
\end{subequations}
only involve $\pBI(\nB)$ and the rate constants prescribed by the reaction scheme, the FLEX approximation provides a unique solution for the conditional probabilities $\pBI(\nB)$ at the interface given the mesoscopic boundary conditions $\rhoBI^{\pm}$.
Moreover, because we ignore fluxes in the condensed phase for the conditions studied in our simulations, \eqref{eq:flex-ss} implies
\begin{align}
  \label{eq:flex-ss-vonly}
  0 &= \jB^+(\nB) + \jI^+(\nB) \\
  \label{eq:flex-ss-j-s}
  \jB^+(\nB) &= \sIB(\nB) - \sBI(\nB).
\end{align}
Note that in the main text, we refer to $\jB^+(\nB)$ as $\jB^{\text{FLEX}}(\nB)$.

The interfacial boundary conditions (BCs) for determining $\rhoBI^\pm$ are as follows:
\begin{enumerate}[label={BC ({\arabic*}).},left=\parindent]
\item $\rhoB^- = \rhoB^{\text{eq,l}}$: \vskip0.5ex
  The condensed-phase B-state concentration is assumed to be unperturbed relative to equilibrium for the parameter regime that we study in simulations, i.e., the strongly phase-separated regime.
\item $\langle \jB^-(\nB) \rangle \equiv \sum_{\nB \in \{1,2,3\}} p(\nB) \jB^-(\nB) = -(\Lambda/4) \nabla \rhoB \big|_{z=0-} = 0$: \vskip0.5ex
  In the strongly phase-separated regime, we assume that fluxes in the condensed phase are negligible.
  This condition fixes $\rhoBl = \rhoB^{\text{eq,l}}$ in conjunction with BC (1), which in turn fixes $\rhoIl = \rhoI^- = \rhoB^{\text{eq,l}} / \Kl$.
  In this expression and henceforth, the notation $\langle \cdot \rangle$ indicates an average over $\nB$ on the interface weighted by $p(\nB)$.
\item $\langle \jB^+(\nB) \rangle \equiv \sum_{\nB \in \{1,2,3\}} p(\nB) \jB^+(\nB) = -(\Lambda/4) \nabla \rhoB \big|_{z=0+}$: \vskip0.5ex
  To ensure mass conservation everywhere, we must match the diffusive flux between the interface and the dilute phase.
  In keeping with the spirit of FLEX, we use the marginal distribution $p(\nB)$ obtained from an equilibrium simulation for this calculation.
  According to \eqref{eq:conc-fields}, the mesoscopic flux on the right-hand side is
  \begin{equation}\label{eq:diffusive_flux_mesoscopic}
    \jB(z = 0+) = \frac{D}{\xi} \left[ \frac{1}{1 + \Kv} \rhoB^+ - \frac{\Kv}{1 + \Kv} \rhoI^+ \right]
  \end{equation}
  for a macroscopically flat interface.
\item $\ln [\langle\pB\rangle / (1 - \langle\pB\rangle)] = 0$: \vskip0.5ex
  This equation states that the nonequilibrium driving force for growth of the condensed phase, which is well approximated by the left-hand side of this equation~\cite{cho2023nucleation}, must be zero in order for the phases to be at coexistence at a nonequilibrium steady state.
  The application of this expression to nonequilibrium conditions is motivated by the particle--hole symmetry of the lattice gas~\cite{cho2023nucleation}.
  As shown in Ref.~\cite{cho2023interface}, this condition can be interpreted as mechanical balance, which is equivalent to the condition of equal osmotic pressure at equilibrium.
  This BC also implies $\langle\pB\rangle = \nicefrac{1}{2}$ for a macroscopically flat interface, which is the definition of the Gibbs dividing surface at coexistence.
\end{enumerate}

The predictions of this theory are compared with simulation results in Figs.~4 and 5 in the main text and in Supplementary \figsref{fig:SI_fix_df} and \ref{fig:SI_linear_kinetics}.
In principle, the FLEX framework can be extended to describe nonequilibrium phase coexistence in which mesoscopic fluxes in the condensed phase are not negligible.
In this case, BC (1) would be removed, BC (2) would be straightforwardly modified to allow for nonzero fluxes in the condensed phase, and BCs (3) and (4) would remain unchanged.
A new fourth BC would then be needed to fully determine the conditions for nonequilibrium phase coexistence.
We leave this question to future work, as additional simulation results would be needed to test any proposals.

\subsection*{Nonequilibrium Interfacial Fluctuations are Required for Mesoscopic Fluxes}

To study fluctuations on the interface, we define an effective internal energy difference between the B and I states for each local environment according to the relation
\begin{equation}
  \label{eq:Df-def}
  \pB(\nB) = \pI(\nB) e^{-\beta\Dfeff(\nB) - \beta\epsilon\nB}.
\end{equation}
If the fluctuations are Boltzmann-distributed, then $\Dfeff(\nB)$ must be the same for all values of $\nB$.

Using the definition in \eqref{eq:Df-def}, we can rearrange \eqref{eq:FLEX-fluxes} and use the steady-state condition, \eqref{eq:flex-ss-vonly}, to write
\begin{equation}
  \label{eq:boltz}
  \left(e^{-\beta\Dfeff(\nB) + \beta\Dfeff^+} - 1\right) \left(\frac{\pI(\nB)}{\rhoI^+}\right) (1 - \rhoB^+ - \rhoI^+) = \left(\frac{1}{\rhoB^+} + \frac{1}{\rhoI^+}\right) \left(\frac{\jB^+(\nB)}{D}\right) \,\forall \nB,
\end{equation}
where $\Dfeff^+$ is defined equivalently to \eqref{eq:Df-def} for $\rhoB^+$ and $\rhoI^+$ with $\nB = 0$.
Now let us assume that the fluctuations are Boltzmann-distributed.
If there are $m$ possible values of $\nB$ on the interface, then we obtain $m - 1$ additional constraints.
Therefore the system of equations for $\rhoBI^\pm$ and $p_\text{B/I}(\nB)$ is now overdetermined, and as a result there is no solution for arbitrary parameter choices.
Hence, the fluctuations cannot be Boltzmann-distributed in general.
To find the conditions under which we do have Boltzmann-distributed fluctuations, we need \eqref{eq:boltz} to be trivially satisfied for all $\nB$, which can only occur if $\Dfeff(\nB) = \Dfeff^+$ for all $\nB$.
This condition then implies that $\jB^+(\nB)/D = 0$ for all $\nB$, and consequently that the mesoscopic quantity $\jB(z=0+)/D = 0$.
Moreover, meeting these conditions requires that the right-hand side of \eqref{eq:flex-ss-j-s} is zero, which means
\begin{equation}\label{eq:reactive_flux_zero}
  \pB(\nB)\kBI(\nB) = \pI(\nB)\kIB(\nB) \,\forall \nB.
\end{equation}
Using \eqref{eq:Df-def}, this relation becomes
\begin{equation}\label{eq:local-detailed-balance}
  \frac{\kBI(\nB)}{\kIB(\nB)} = e^{\beta\Dfeff + \beta\epsilon \nB} \,\forall \nB.
\end{equation}
\eqref{eq:local-detailed-balance} is the condition for local detailed balance with $\nB$-independent backward rate constants $\kIB(\nB) = \text{const.}$, which we refer to as a homogeneous reaction scheme~\cite{cho2023nucleation}.
Because this equation must also hold for $r > R$, where we have assumed that $\nB = 0$ in the dilute phase, we can write $\Dfeff^+$ explicitly as
\begin{equation}
  \label{eq:Dfeff-homo}
  \beta\Dfeff^+ = \beta\Df + \ln \left[ \frac{\kIB^{\text{p}}(\nB = 0) + \kIB^{\text{d}}(\nB = 0) e^{\beta\Delta\mu}}{\kIB^{\text{p}}(\nB = 0) + \kIB^{\text{d}}(\nB = 0)} \right].
\end{equation}
In summary, achieving Boltzmann statistics for the conditional probabilities on the interface requires that $\jB(z=0+)/D = 0$, which in turn requires that $\kIB$ is independent of $\nB$.
If we instead assume that mesoscopic fluxes are present in the dilute phase with a finite self-diffusion coefficient $D$, which requires $\jB(\nB) \ne 0$ for some $\nB$ in \eqref{eq:boltz}, then we must have $\Dfeff(\nB) \ne \Dfeff^+$, which means that the statistics cannot be Boltzmann throughout the system and $\kIB$ must have a non-constant $\nB$-dependence.

In the case of a homogeneous reaction scheme, which may correspond, for example, to a scenario in which the enzymes that are required for the driven reaction pathway are homogeneously distributed in the condensed and dilute phases~\cite{kirschbaum2021controlling,cho2023nucleation}, the reactions must still satisfy the local detailed balance constraint given by Eq.~(2) in the main text.
The driven reaction rate constants in this scenario are thus
\begin{subequations}
  \label{eq:homo-rates}
\begin{align}
    &\kIB^\text{d}(\nB) = \eta\\
    &\kBI^\text{d}(\nB) = \eta\exp(\beta\Df + \beta\Dmu + \beta\epsilon\nB),
\end{align}
\end{subequations}
while the passive reaction rate constants remain unchanged, so that \eqref{eq:Dfeff-homo} reduces to
\begin{equation}\label{eq:effective_Df}
    \beta\Dfeff^+ = \beta\Df + \ln\left[\dfrac{1+\eta e^{\beta\Dmu}}{1+\eta}\right].
\end{equation}
Simulations confirm the prediction that the reaction scheme given by \eqref{eq:homo-rates} does not result in mesoscopic fluxes, even when the system is driven far from equilibrium (\figref{fig:SI_homogeneous}).
This concentration-independent reaction scheme dissipates energy at a higher rate in the dilute phase (\figref{fig:SI_homogeneous}A), but nonetheless leads to zero net mesoscopic diffusive flux (\figref{fig:SI_homogeneous}B).
This result demonstrates that spatially inhomogeneous entropy production due to microscopic transitions occurring in coexisting phases does not necessarily guarantee observable nonequilibrium effects at mesoscopic scales~\cite{Seifert2019Inference}.
Rather, the emergence of mesoscopic nonequilibrium effects depends on the microscopic details of the system and, according to our theory, is mediated by fluctuations at the phase-separated interface.

\subsection*{Limiting Behaviors in the Diffusion-dominant and Reaction-dominant Regimes}
\label{sec:limiting_behaviors}
Elements of the effective equilibrium analysis introduced in the previous section are also useful for investigating the limiting behaviors in the diffusion-dominant ($\Lambda\rightarrow\infty$) and reaction-dominant ($\Lambda\rightarrow 0$) regimes.
In the diffusion-dominant limit, the self-diffusion coefficient $D = \Lambda/4$ diverges.
However, the diffusive flux $\jB(\nB)$ must remain finite for all values of $\nB$ to satisfy \eqref{eq:flex-ss-j-s} within the FLEX framework for finite $\beta\Dmu$ and physical values of $\pBI(\nB)$.
Therefore, $\jB(\nB)/D$ approaches zero in this limit, and the interfacial statistics follow the Boltzmann distribution as indicated by \eqref{eq:boltz}, with the effective internal energy difference $\Dfeff^+$ in \eqref{eq:Dfeff-homo} governing the distributions in both the dilute phase and on the interface simultaneously.
For this reason, the capillary fluctuations approach the equilibrium distribution in this limit, as shown in Fig.~4D in the main text.
Finiteness of $\jB$ also implies that the concentration gradient $\nabla\rhoB|_{R+}$ associated with the flux in BC~(3) must vanish, leading to a uniform concentration profile in the dilute phase, such that the effective equilibrium description on the interface extends to the far field.

This effective equilibrium behavior in the diffusion-dominant limit can also be understood intuitively by considering the implied separation of timescales.
In the kinetic scheme introduced in Eq.~(1) in the main text, the fundamental time unit is defined by the passive $\IB$ reaction.
However, if time were instead measured with respect to the timescale for diffusive events, then reactions would occur at a vanishingly slow rate in the rescaled time units, which is consistent with the notion of an effective thermal equilibrium.
Viewed this way, it is clear that the behavior in the diffusion-dominant regime is consistent with the assumptions of linear irreversible thermodynamics~\cite{de2013non}, which is valid for small perturbations from equilibrium (or for small perturbations from an effective equilibrium in which the system is globally described by Boltzmann statistics with a common $\Dfeff$).
This observation explains why our nonequilibrium sharp-interface theory and the alternative theory shown by the dotted lines in Fig.~4A in the main text (see below for details) match in the diffusion-dominant limit, since the latter assumes that all thermodynamic potentials and interfacial material properties are equivalent to the equilibrium system.

By contrast, in the reaction-dominant regime, the statistics of interfacial fluctuations do not necessarily follow the Boltzmann distribution, and the system cannot be described by a global effective equilibrium with a common value of $\Dfeff$ when the reaction scheme involves concentration-dependent rate constants.
In this case, $\jB(\nB)$ approaches zero due to the vanishing self-diffusion coefficient.
This eliminates the net reactive flux in \eqref{eq:flex-ss-j-s} so that the interfacial distributions $\pBI(\nB)$ must be given by \eqref{eq:reactive_flux_zero}.
In \eqref{eq:reactive_flux_zero}, any dependence of $\kIB(\nB)$ on the local concentrations leads to deviations from Boltzmann statistics, since the ratio $\pB(\nB)/\pI(\nB)$ is no longer proportional to $\exp(-\beta\Dfeff-\beta\epsilon\nB)$ unless the reaction scheme is homogeneous.
For these reasons, the capillary fluctuations deviate from equilibrium in the reaction-dominant limit, as shown in Fig.~4D in the main text.
Nonzero finite values of $\jB^+(\nB)/D$ indicate that the mesoscopic quantity $\jB^+/D$ also remains finite in this limit.
However, the mesoscopic fluxes vanish, as shown by $J_{\text{B}}/\xi$ in Fig.~4C in the main text, as does the dilute-phase concentration gradient for $r > R$.
We note that this limit is closely related to the non-conservative simulation scheme studied in Refs.~\cite{cho2023nucleation} and \cite{cho2023interface}.

\subsection*{Nonequilibrium Effects are Maximized when Fluxes Through the Passive and Driven Reaction Pathways are Comparable}

\begin{figure}
    \centering
    \includegraphics[width=0.95\textwidth]{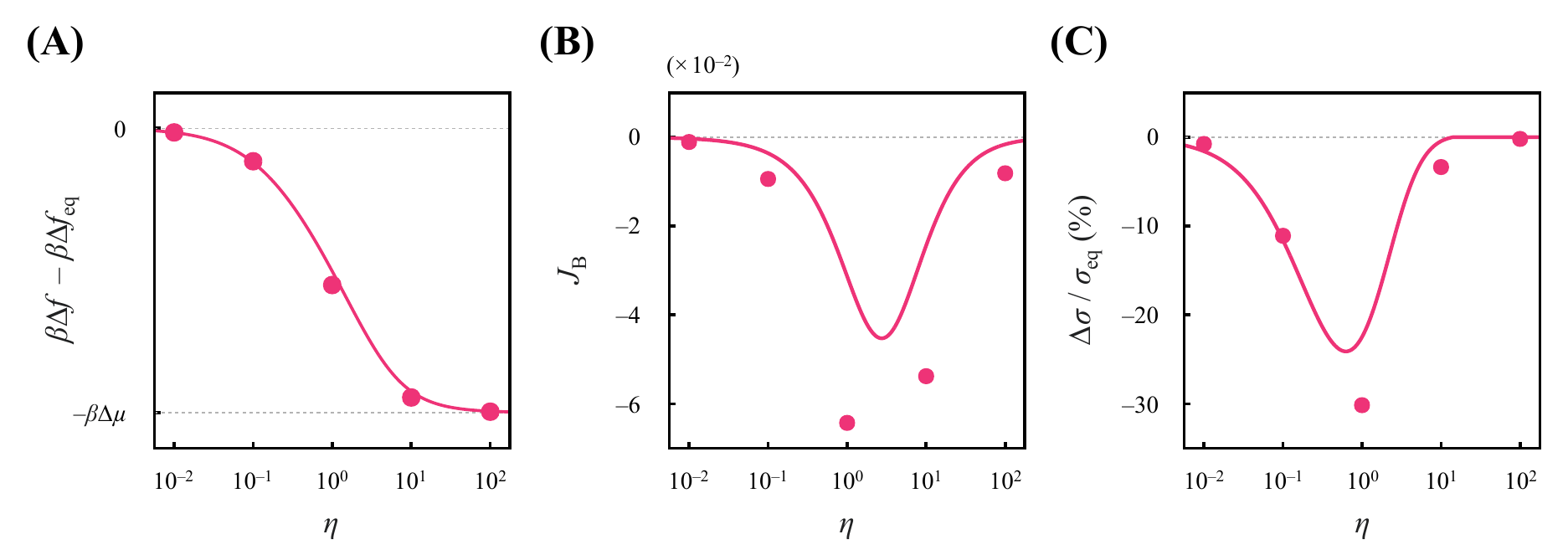}
    \caption{\textbf{Mesoscopic nonequilibrium effects are maximized when the reactive fluxes along the driven and passive reaction pathways are comparable.}
    (A)~The dependence of the constant-$\rhov$ coexistence curve on the relative flux along the passive and driven reaction pathways, ranging from the purely passive limit $(\eta\rightarrow0)$ to the purely driven limit $(\eta\rightarrow\infty)$.
    (B)~The integrated mesoscopic diffusive flux density for bonding-state molecules for a macroscopically flat interface, $\JB=\int_0^\infty \jB\,dz$.
    (C)~The deviation in the interfacial tension, $\Delta\sigma\equiv \sigma - \sigma_\text{eq}$, from the equilibrium value, $\sigma_\text{eq}$, along the coexistence curve in panel~A.
    The values of $\sigma$ are determined by fitting our theory to the simulation measurements of the far-field dilute-phase total concentration as a function of the droplet radius, $R$, as shown in Fig.~5 in the main text and described in the \textit{Materials and Methods}.
    The predicted values of $\sigma$ are computed by evaluating an equilibrium expression for the interfacial tension~\cite{shneidman1999analytical} using the effective interaction strength predicted by \eqref{eq:effective_interaction_strength}.
    In all panels, marks indicate simulation results, and the solid lines show theoretical predictions.
    All data and predictions presented are obtained for the conditions $\beta\Dmu = 2$ and $\Lambda = 10^2$, with $\beta\Df$ tuned such that $\rhovf = 0.05$, as in Fig.~4 in the main text.}
    \label{fig:SI_eta}
\end{figure}

Nonequilibrium effects due to the chemical drive vanish in both the passive-reaction dominant $(\eta\rightarrow0)$ and the driven-reaction dominant $(\eta\rightarrow\infty)$ limits.
In the former limit, the steady state reaches thermal equilibrium with the internal energy difference $\Df$.
In the latter limit, the steady state distribution is equivalent to an equilibrium distribution with $\Df\rightarrow\Df-\Dmu$, since both the forward and backward reactions along the dominant pathway are modified by $\Dmu$ in accordance with the local detailed balance condition.
These limiting behaviors are predicted for the homogeneous reaction case by \eqref{eq:effective_Df} and are consistent with the phase diagrams determined for the inhomogeneous reaction scheme with concentration-dependent rate constants that is studied in the main text (\figref{fig:SI_eta}A).

By contrast, nonequilibrium fluctuations are maximized when the reactive fluxes along the two pathways are comparable.
In \figref{fig:SI_eta}B,C, we show that both the nonequilibrium diffusive flux densities and the deviation of the interfacial tension from the equilibrium value are most pronounced near $\eta \approx 1$ for the inhomogeneous reaction scheme studied in the main text.

\subsection*{Predicting Nonequilibrium Interfacial Height Fluctuations}

Although the FLEX framework assumes that the marginal distribution of local environments, $p(\nB)$, is unperturbed by the nonequilibrium chemical drive, we can nonetheless predict qualitative changes in the capillary fluctuations by following the effective-equilibrium approach introduced in Refs.~\cite{cho2023nucleation,cho2023interface}.
Specifically, we apply equilibrium capillary fluctuation theory by considering an effective ``single-bond'' interaction strength between a microscopically flat interface and a single adsorbed B-state molecule.
From the conditional distribution $\pBI(\nB)$ at phase coexistence, we compute the effective interaction
\begin{equation}\label{eq:effective_interaction_strength}
    \beta\tilde{\epsilon} \equiv \ln\left[\dfrac{\pB(\nB=1)}{1-\pB(\nB = 1)}\right],
\end{equation}
from which we can predict the associated interface height fluctuations based on an equilibrium solid-on-solid model~\cite{saito1996statistical},
\begin{equation}\label{eq:effective_height_fluctuation}
    \dfrac{\langle \Dh^2 \rangle}{\langle \Dh^2 \rangle}_\text{eq} = \dfrac{\sinh^2(\beta\epsilon/4)}{\sinh^2(\beta\tilde{\epsilon}/4)}.
\end{equation}

We emphasize that this approach to predicting nonequilibrium height fluctuations is highly approximate, since we only consider a single local configuration ($\nB = 1$) and because the FLEX approximation ignores coupling between chemical reactions and collective rearrangements of the interface.
Although our approach captures qualitative trends, the predictions are generally not quantitatively accurate, as shown in Fig.~4D in the main text.
Inaccuracies are particularly pronounced when the mobility $\Lambda$ is large, since the approximations underlying FLEX are less applicable in this regime.
Moreover, our approach may predict $\beta\tilde\epsilon = \beta\epsilon$, and thus $\langle \Dh^2 \rangle = \langle \Dh^2 \rangle_{\text{eq}}$, even when there are non-Boltzmann fluctuations at the interface, which is not consistent with the simulation results shown in Fig.~4D in the main text.
This inaccuracy occurs because the effective bond strength predicted by the FLEX approximation with $\nB = 1$, \eqref{eq:effective_interaction_strength}, may be extremely similar to the actual bond strength even when the conditional distribution at the interface does not follow Boltzmann statistics overall.
Finally, because FLEX does not predict changes in the marginal distribution $p(\nB)$, this approach does not predict changes in the $q$-dependent scaling of the fluctuation spectrum that are observed in simulations when $\langle \Dh^2 \rangle \gg \langle \Dh^2 \rangle_{\text{eq}}$ (\figref{fig:SI_capillary}).
Nonetheless, the qualitative predictions of this approach are consistent with the formalism developed above in the sense that this approach always predicts $\Dh = \Dh_\text{eq}$ when the interfacial statistics follow the Boltzmann distribution.
Moreover, it predicts the correct limiting behaviors in the reaction-dominant and diffusion-dominant regimes.

\subsection*{Application to Finite-size Droplets with Curved Interfaces}

To apply this theory to a droplet with finite radius $R$, we modify the last two interfacial boundary conditions as follows:
\begin{enumerate}[label={BC ({\arabic*}).},left=\parindent,start=3]
\item $\langle \jB^+(\nB) \rangle \equiv \sum_{\nB \in \{1,2,3\}} p(\nB) \jB^+(\nB) = -(\Lambda/4) \nabla \rhoB \big|_{R+}$: \vskip0.5ex
  According to \eqref{eq:diffusive_flux_density}, the mesoscopic flux on the right-hand side is now
  \begin{equation} \label{eq:diffusive_flux_density_curved}
    \jB(r = R+) = \frac{D}{\xi} \left[ \frac{1}{1 + \Kv} \rhoB^+ - \frac{\Kv}{1 + \Kv} \rhoI^+ \right] \frac{K_1(\xi^{-1}R)}{K_0(\xi^{-1}R)}.
  \end{equation}
\item $\ln [\langle\pB\rangle / (1 - \langle\pB\rangle)] - \beta\sigma/R = 0$: \vskip0.5ex
  This equation relates the nonequilibrium driving force for the growth of the condensed phase to a phenomenological Gibbs--Thomson correction with nonequilibrium interfacial tension $\sigma$.
  At equilibrium, the first term represents the osmotic pressure difference between the two phases, so that this BC is equivalent to the Young--Laplace equation for a two-dimensional fluid, $\beta\Delta P = \beta\sigma_{\text{eq}}/R$, with the equilibrium interfacial tension $\sigma_\text{eq}$~\cite{porter2021phasetransformation}.
  Here we extend this concept to nonequilibrium conditions, whereby the first term represents the nonequilibrium potential difference across the interface~\cite{cho2023interface} and the second term assumes that a similar Gibbs--Thomson-like correction still plays a role, albeit with a nonequilibrium interfacial tension, leading to mechanical balance across the curved interface at phase coexistence.
  This application of a Gibbs--Thomson-like correction is consistent with prior simulations of chemically driven droplet nucleation~\cite{cho2023nucleation}, which can be described quantitatively by assuming a $\beta\sigma/R$ term in the expression for the nonequilibrium nucleation barrier.
\end{enumerate}

At equilibrium, BC~(3) does not affect the size scaling of droplets due to the absence of diffusive fluxes.
Yet under nonequilibrium conditions, the droplet curvature appears in both BC~(3) via \eqref{eq:diffusive_flux_density_curved} and explicitly in BC~(4).
Whether the droplet follows an equilibrium-like size scaling relation therefore depends on the relative size of the droplet compared to the reaction--diffusion length in the dilute phase.

We first consider the scenario in which the reaction--diffusion length is small compared to the droplet radius, $\xi \ll R$.
In this case, the Bessel function term in \eqref{eq:diffusive_flux_density_curved} remains close to unity (\figref{fig:SI_Bessel}), such that BC~(3) becomes insensitive to $R$.
Qualitatively, this means that the interface appears to be locally flat on length scales comparable to $\xi$.
The droplet-size dependence of the coexistence conditions is then entirely contained within BC~(4) so that we may expect an equilibrium-like scaling of the form $\ln(\rhov / \rhovf) \sim R^{-1}$, which matches the simulation observations in Fig.~5A in the main text.

In the opposite scenario where the reaction--diffusion length is comparable to or greater than the droplet radius, $\xi \gtrsim R$, the scaling relation can be more complicated due to the $R$-dependence of the nonequilibrium diffusive flux.
In this case, the Bessel function term in \eqref{eq:diffusive_flux_density_curved} is sensitive to $R$ (\figref{fig:SI_Bessel}), so that the size scaling is governed by both boundary conditions.
Consequently, $\ln(\rhov / \rhovf)$ can become a nonlinear function of $R^{-1}$, as shown in Fig.~5B in the main text.
Nonetheless, our theory predicts that nonequilibrium droplets in both the reaction-dominant $(\Lambda\rightarrow0)$ and the diffusion-dominant $(\Lambda\rightarrow\infty)$ limits revert to the equilibrium-like scaling form $\ln(\rhov / \rhovf) \sim R^{-1}$.
In the reaction-dominant regime, the diffusive flux vanishes so that the droplet-size dependence only enters BC~(4), as in equilibrium; however, the interfacial tension may differ from equilibrium since the interfacial fluctuations may not follow Boltzmann statistics in this limit.
On the other hand, in the diffusion-dominant limit, both the scaling form and the interfacial tension are the same as in equilibrium, since the system approaches a global effective equilibrium due to the vanishing concentration gradient in the dilute phase.

We emphasize that the effect of nonequilibrium fluxes in BC~(3) is separate from the change in the interfacial tension and that both effects are required to explain the observed size scaling under nonequilibrium conditions.
For demonstration purposes, we show the predictions of an alternative theory in which the interfacial tension in BC~(4) is fixed to the equilibrium value, $\sigma_{\text{eq}}$, in \figref{fig:SI_sigma}.
This alternative theory accurately predicts nonlinear size scaling for intermediate values of $\Lambda$ but does not agree quantitatively with the simulation data due to the neglect of the nonequilibrium interfacial tension.

\subsection*{Alternative Sharp-interface Theory with Equilibrium Interfacial Boundary Conditions}

Existing sharp-interface theories, including Ref.~\cite{wurtz2018chemical} and Ref.~\cite{bauermann2022energy}, assume that equilibrium phase-coexistence conditions apply at the sharp interface in the presence of nonequilibrium mesoscopic fluxes.
Here we follow the approach taken in Ref.~\cite{bauermann2022energy}, although other theories that assume equilibrium phase coexistence at the interface, such as Ref.~\cite{wurtz2018chemical}, lead to similar conclusions.

In this approach, it is first necessary to propose an equilibrium model for the free energy of the system.
To this end, we utilize a ternary regular solution model~\cite{porter2021phasetransformation} to compute the chemical potentials of the two solute species, $\muB$ and $\muI$, and the osmotic pressure, $P$, as a function of $\rhoB$ and $\rhoI$:
\begin{subequations}
  \label{eq:rs}
  \begin{align}
    \beta\muB(\rhoB, \rhoI) &= \ln \rhoB - \ln(1 - \rhoB - \rhoI) + 4\epsilon\rhoB \\
    \beta\muI(\rhoB, \rhoI) &= \ln \rhoI - \ln(1 - \rhoB - \rhoI) \\
    \beta P(\rhoB, \rhoI) &= -\ln (1 - \rhoB - \rhoI) + 2 \epsilon \rhoB^2.
  \end{align}
\end{subequations}
These formulas are accurate for lattice models when applied far from a critical point and are thus appropriate for modeling the strongly phase-separated regime that we consider in our simulations.

By contrast to our theoretical approach, an equilibrium model of the sharp interface implies that no reactions take place on the interface itself.
Maintaining mass conservation therefore requires that we consider mesoscopic fluxes in both phases.
Specifically, conservation of molecules at a macroscopically flat interface requires
\begin{equation} \label{eq:alternative_mass_balance}
  \jBI(z=0-) = \jBI(z=0+),
\end{equation}
which, by applying \eqref{eq:conc-fields}, leads to 
\begin{equation}
  -\dfrac{\rho_{\text{S}}^{\text{l}}}{\xil}\left[\dfrac{1}{\zeta+\Kl}\rhoB^{-}-\dfrac{\Kl}{\zeta+\Kl}\rhoI^{-}\right] = \dfrac{1}{\xi}\left[\dfrac{1}{1+\Kv}\rhoB^{+}-\dfrac{\Kv}{1+\Kv}\rhoI^{+}\right].
  \label{eq:conserv}
\end{equation}
In the strongly phase-separated limit, the reaction--diffusion lengths in the dilute and condensed phases are
\begin{equation}
  \xi %%&= \sqrt{[\kBI^{\circ,\text{v}}/D + \kIB^{\circ,\text{v}}/D} = \sqrt{D/(\kBI^{\circ,\text{v}}+\kIB^{\circ,\text{v}})}
  = \sqrt{\Lambda/4[\kBI(\nB=0)+\kIB(\nB=0)]}
  \qquad \text{and} \qquad 
  \xil %%&= \sqrt{[\kBI^{\circ,\text{l}}/\DBl + \kIB^{\circ,\text{l}}/\DIl} = \sqrt{\DBl/(\kBI^{\circ,\text{l}}+\zeta^{-1}\kIB^{\circ,\text{l}})}
  = \sqrt{\rho_{\text{S}}^{\text{l}} \zeta^{-1} \Lambda/4[\kBI(\nB=4)+\zeta^{-1}\kIB(\nB=4)]},
\end{equation}
respectively.
Importantly, the expressions for $\xi$ and $\xil$ imply that the mobility $\Lambda$ drops out of \eqref{eq:conserv}.
For this reason, the phase-coexistence predictions of this theory are independent of $\Lambda$ in Fig.~4A in the main text and in Supplementary \figref{fig:SI_linear_kinetics}B.

Following Ref.~\cite{bauermann2022energy}, we then define four interfacial boundary conditions to uniquely determine $\rhoBI^\pm$:
\begin{enumerate}[label={BC ({\arabic*}).},left=\parindent]
\item $\muB(\rhoB^+,\rhoI^+) = \muB(\rhoB^-,\rhoI^-)$: Equal chemical potentials for the B-state molecule in the coexisting phases,
\item $\muI(\rhoB^+,\rhoI^+) = \muI(\rhoB^-,\rhoI^-)$: Equal chemical potentials for the I-state molecule in the coexisting phases,
\item $P(\rhoB^+,\rhoI^+) = P(\rhoB^-,\rhoI^-)$: Equal osmotic pressures in the coexisting phases, and 
\item \eqref{eq:conserv}: Mass conservation at the interface.
\end{enumerate}

The predictions of this theory for the case of a macroscopically flat interface are shown with dashed lines in Fig.~4A in the main text and in Supplementary \figref{fig:SI_linear_kinetics}B.
As noted above, the nonequilibrium coexistence curve predicted by this theory is always independent of $\Lambda$ since $\jBI\propto \Lambda$ on both sides of the interface $(z=0\pm)$, leading to exact cancellation of the $\Lambda$ dependence on the left and right-hand sides of \eqref{eq:alternative_mass_balance}.
Furthermore, if we compute phase coexistence as a function of $\Df$ while holding the total far-field concentration $\rhov \equiv \rhoBv + \rhoIv$ constant, then we find that the values of $\rhoBI^\pm$ predicted by this theory are also independent of $\Delta\mu$.
This is because the ideal-gas assumption for the dilute phase, which is used in the derivation of \eqref{eq:conserv}, leads to $\rhoB^+ + \rhoI^- = \text{const.}$ under this constraint.
Then, because the equilibrium conditions in BCs (1)--(3) define a one-dimensional binodal in the $\rhoB$--$\rhoI$ plane, this constraint on $\rhoBI^+$ always picks out the same coexistence point and tie line.
Consequently, we can solve for coexistence under the constraint $\rhov = \text{const.}$ simply by evaluating \eqref{eq:coex} with $\rhoBI^\pm$ set equal to their values when $\Delta\mu = 0$ and $\Df = \Df_{\text{eq}}$.
As shown in Fig.~4A in the main text, these predictions only agree with the simulation results in the limit $\Lambda\rightarrow\infty$, since the system approaches a global effective equilibrium in this regime.

Finally, we note that this theory predicts that the microscopic and mesoscopic properties of the phase-separated interface remain unperturbed relative to equilibrium by construction.
Consequently, this theory predicts that the conditional distribution $\pBI(\nB)$ always follows Boltzmann statistics, which is implicitly assumed in the definitions of $\muB$, $\muI$, and $P$ in \eqref{eq:rs}, and that the capillary fluctuations are unchanged relative to equilibrium.
This fundamental feature of the theory is qualitatively at odds with the simulation results shown in Fig.~3 and Fig.~4D in the main text.
Moreover, this theory does not predict a nonequilibrium effective interfacial tension due to the chemical drive.
Therefore, this theory is also unable to predict the changes in the $R$-dependent far-field concentrations that are presented in Fig.~5 in the main text.

\clearpage

 \clearpage

\begin{figure}
    \centering
    \includegraphics[width=\linewidth]{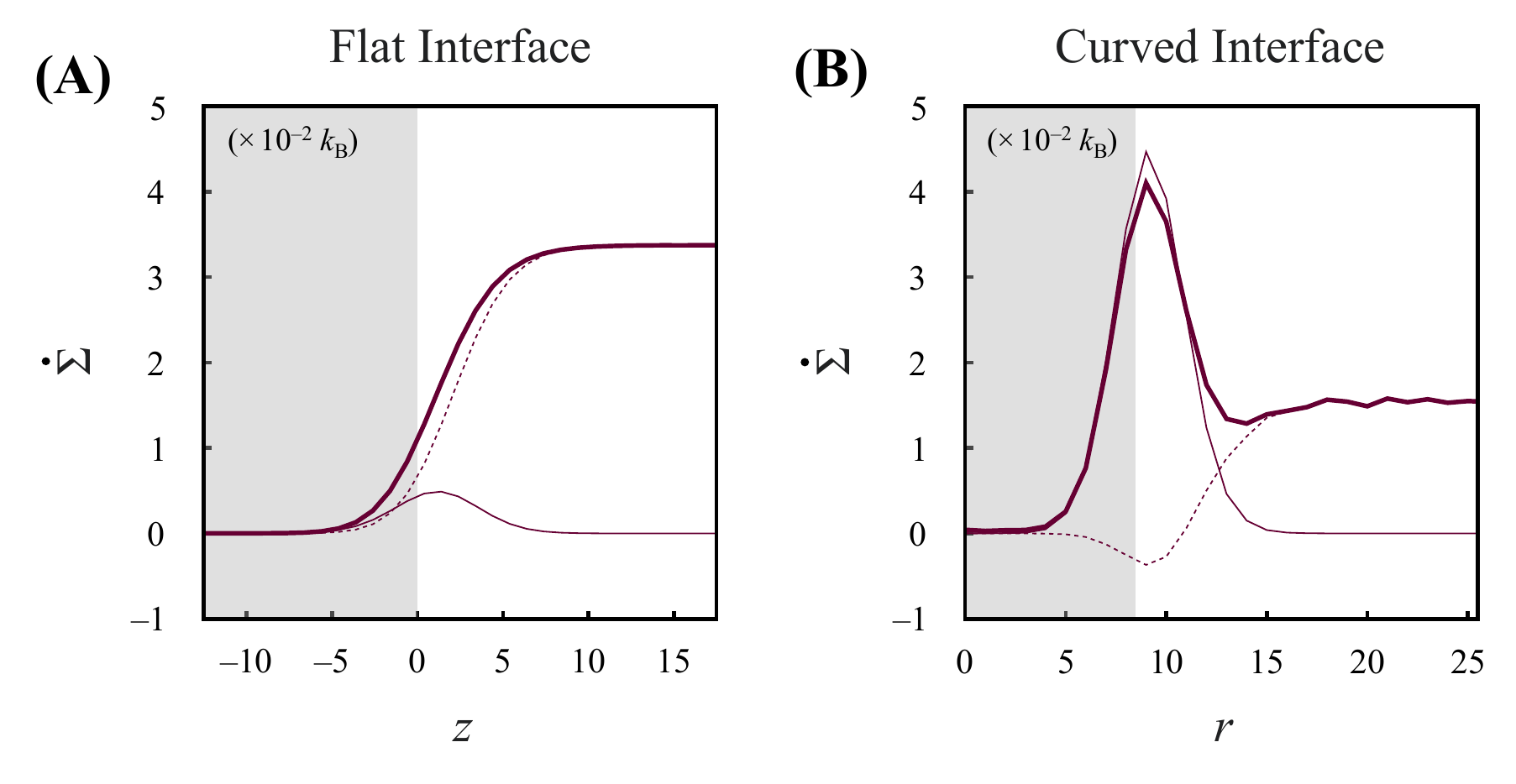}
    \caption{\textbf{Entropy production rate density profiles at nonequilibrium phase coexistence.}
    The entropy production rate density, $\dot{\Sigma}$, at coexistence between the condensed (shaded) and dilute (unshaded) phases in simulations with (A) macroscopically flat and (B) curved interfaces.
    Thick solid lines show the total entropy production rate, whereas thin solid and dotted lines represent the contributions from the interface and the bulk dilute phase, respectively.
    The contribution from the bulk condensed phase is negligible by comparison and is thus omitted.
    Note that panels A and B correspond to the left and right columns of Fig.~2 in the main text, respectively, and are cross-sections of the spatial maps shown in Fig.~1E,F in the main text.
    }
    \label{fig:SI_entropy_profile}
\end{figure}

\begin{figure}
    \centering
    \includegraphics[width=0.9\textwidth]{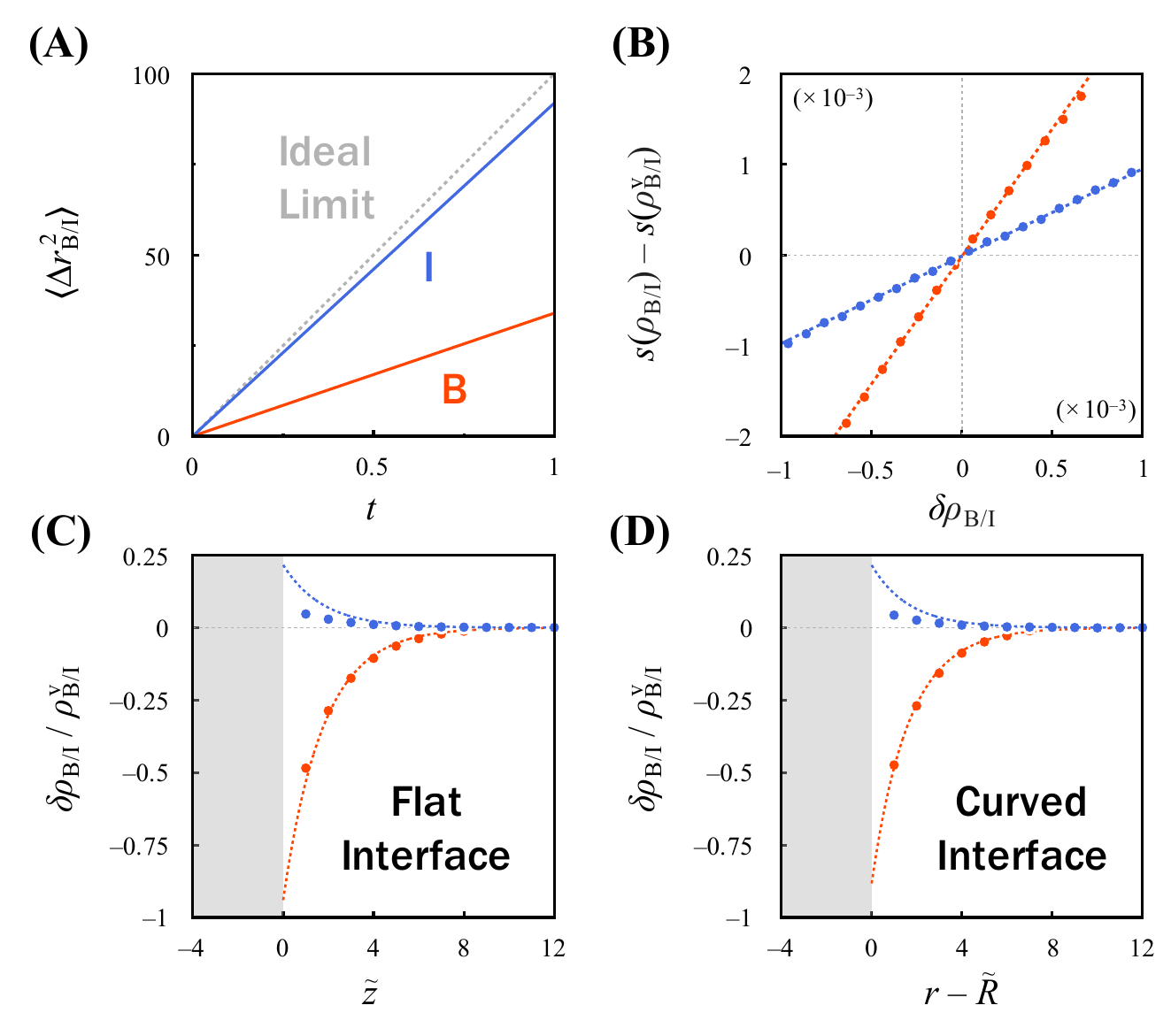}
    \caption{\textbf{Test of the linearized reaction--diffusion theory in the dilute phase.}
    (A)~The mean-squared displacement, $\langle \Delta r^2 \rangle$, linearly increases with time, $t$, for both the B-state (orange) and I-state (blue) molecules, albeit more slowly than in the ideal limit, $\langle \Delta r^2 \rangle = \Lambda t$ (gray dotted line).
    (B)~The linearized reaction rates, $\kTBI$ and $\kTIB$, are fit according to the linear slope (dotted lines) of the change in the reactive flux from the average value, $s(\rhoBI) - s(\rhoBI^\text{v})$, with respect to the perturbation in the concentration, $\delta\rhoBI \equiv \rhoBI - \rhoBI^\text{v}$.
    (C,D)~The deviation in the dilute-phase concentration profiles from the far-field values in direct-coexistence simulations of the dilute (unshaded) and condensed (shaded) phases with (C)~macroscopically flat and (D)~curved interfaces.
    The concentration profiles are measured relative to the instantaneous fluctuating interface located at $\tilde{z} = 0$ and $r = \tilde{R}$, respectively.
    Marks indicate simulation results, whereas orange dotted lines indicate fits to \eqsref{eq:rhoB_2D} and (\ref{eq:rhoB_1D}) within the dilute phase using the empirical reaction--diffusion length, $\xi$, determined from panels A and B.
    Blue dotted lines are the prediction of \eqref{eq:local_mass_conservation_dilute_phase}, and are thus identical to the orange dotted lines except for a change of sign.
    For the fit in panel D, the average droplet radius $R = 8.46$ is used in \eqref{eq:rhoB_2D}.
    Extrapolating the fitted B-state concentration profiles to the interface predicts physically reasonable values, such that $\rhoB^+ > 0$ at the interface, in both cases.
    All simulation data are obtained at $\beta\Dmu = 2$, $\Lambda = 10^2$, $\beta\Df = 1.7337$; the far-field concentrations are $\rhov = 0.05$ for panels A--C and $\rhov = 0.0548$ for panel D.
    Note that panels C and D correspond to the left and right columns, respectively, of Fig.~2 in the main text.
    }
    \label{fig:SI_RD_test}
\end{figure}

\begin{figure}
    \centering
    \includegraphics[width=0.75\linewidth]{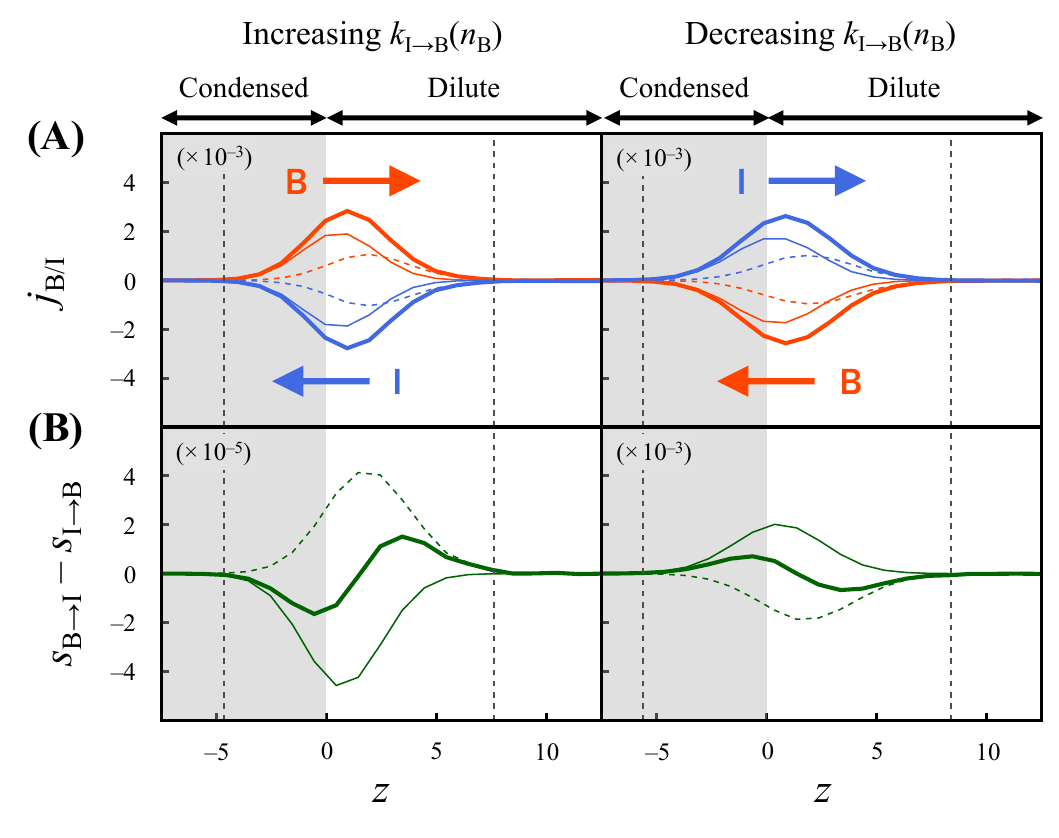}
    \caption{\textbf{The direction of the mesoscopic fluxes can be reversed by altering either the sign of the chemical drive or the concentration-dependence of the reaction rate constants.}
    (A)~The mesoscopic diffusive fluxes $\jB$ (orange) and $\jI$ (blue) and (B) the net mesoscopic reactive flux in the $\BI$ direction, $\sBI - \sIB$, near a macroscopically flat interface between the coexisting condensed (shaded) and dilute phases (unshaded).
    In the left column, the $\IB$ reaction rate constant along the driven reaction pathway increases with respect to the number of nearest-neighbor B-state molecules, $\nB$, as described by Eq.~(1) in the main text, whereas in the right column, the $\IB$ reaction rate constant along the driven reaction pathway decreases with respect to $\nB$ (see the lower panel in \figref{fig:SI_linear_kinetics}A for the reaction scheme).
    The simulation parameters are $\beta\Dmu = -2$ and $\Lambda = 10^2$ for both the left and right columns.
    To compare simulation results at the same total concentration $\rhov = 0.05$ in the far-field dilute phase, we choose $\beta\Df = 3.3150$ for the left column and $\beta\Df = 3.3180$ for the right column.
    Note that the directions of the diffusive fluxes in panel A and the signs of the reactive fluxes in panel B are reversed between the columns.
    Similarly, the properties shown in the left column are reversed relative to Fig.~2B,C in the main text, in which case $\beta\Dmu = 2$.
    The vertical dashed lines mark the boundaries of the mesoscopic flux region where $|\jB|$ is greater than 1\% of the peak value.}
    \label{fig:SI_flux_more}
\end{figure}

\begin{figure}
    \centering
    \includegraphics[width=0.4\linewidth]{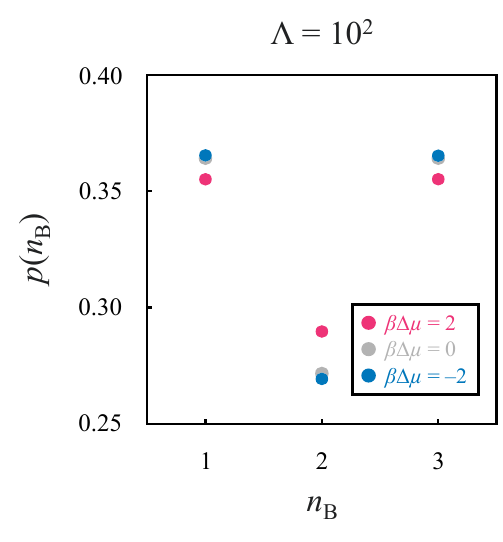}
    \caption{\textbf{The marginal distribution of local environments on the interface deviates from equilibrium.}
    The marginal distribution, $p(\nB)$, on instantaneous interfaces sampled both at equilibrium and under the far-from-equilibrium conditions shown in Fig.~3 in the main text.
    At $\beta\Dmu = 2$ (pink), the populations of $\nB = 1$ and $\nB  = 3$ lattice sites decrease, and the population of $\nB = 2$ lattice sites increases, relative to equilibrium (gray).
    This behavior is consistent with the enhanced capillary-wave amplitudes shown in Fig.~3A in the main text.
    }
    \label{fig:SI_marginal}
\end{figure} \clearpage

\begin{figure}
    \centering
    \includegraphics[width=\linewidth]{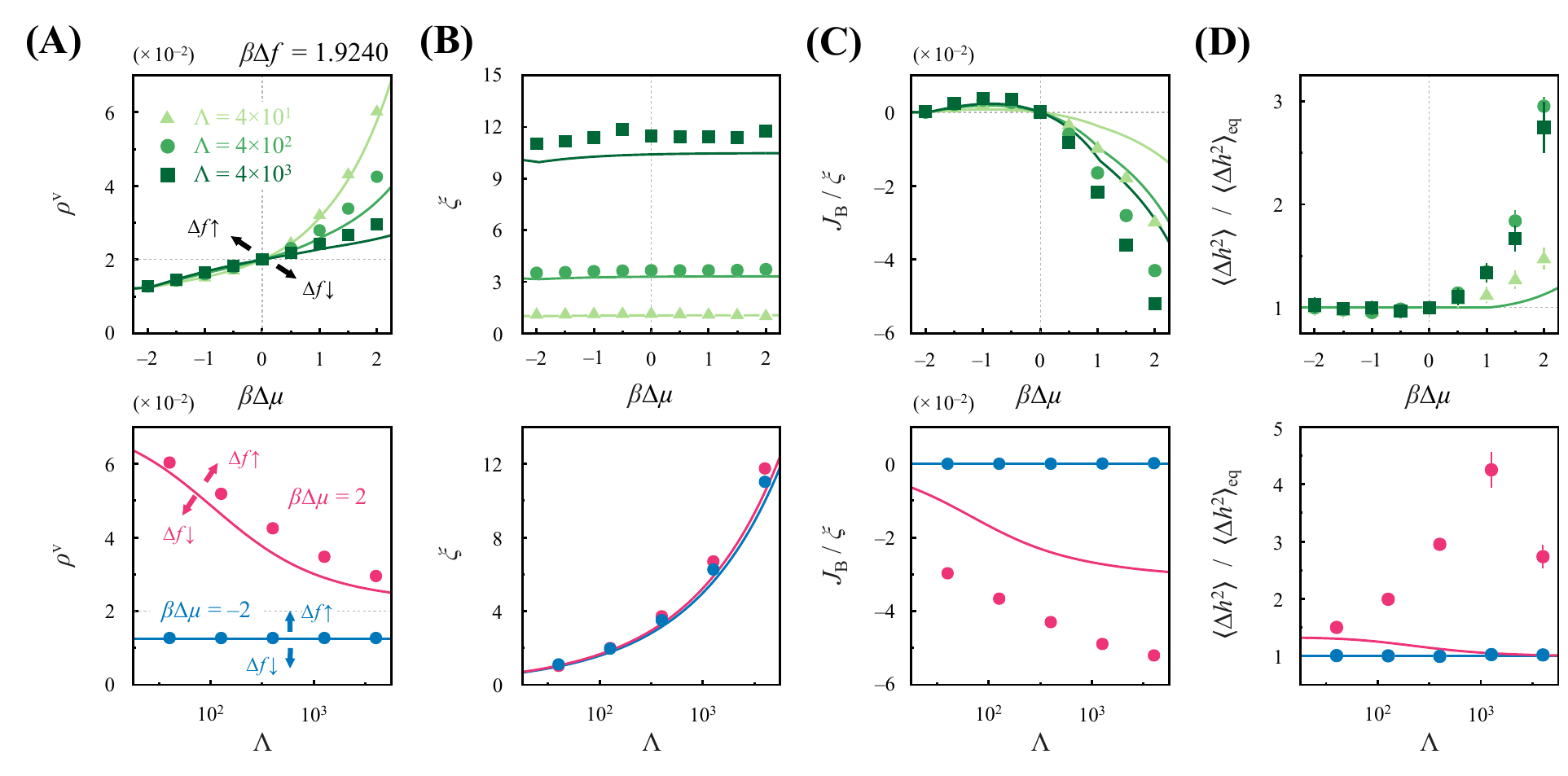}
    \caption{\textbf{Prediction of nonequilibrium phase coexistence with a fixed internal energy difference, $\Df$.}
    Simulation results (marks) and theoretical predictions from the nonequilibrium sharp-interface theory (lines) for a flat interface $(R\rightarrow\infty)$ at constant $\Lambda$ (upper panels) and $\beta\Dmu$ (lower panels).
    (A)~Phase diagram showing the coexistence curves with a fixed internal energy difference, $\Df$.
    Increasing $(\Df\!\uparrow)$ or decreasing $(\Df\!\downarrow)$ the internal energy difference shifts the far-field total concentration in the dilute phase, $\rhov$, as indicated.
    (B)~The reaction--diffusion length in the dilute phase, $\xi$,
    (C)~the spatially averaged diffusive flux of B-state molecules near the interface, $\JB/\xi$, and
    (D)~the change in the mean-squared interfacial height fluctuations, $\langle\Dh^2\rangle$, relative to the equilibrium value, $\langle\Dh^2\rangle_\text{eq}$, at $\Dmu = 0$.
    All data and predictions in panels B--D are made along the coexistence curves shown in panel A.
    The system parameters are $\beta\epsilon = -3.00$ and $\beta\Df = 1.9240$, which corresponds to $\rhov = 0.02$ at equilibrium.}
    \label{fig:SI_fix_df}
\end{figure}

\begin{figure}
    \centering
    \includegraphics[width=\linewidth]{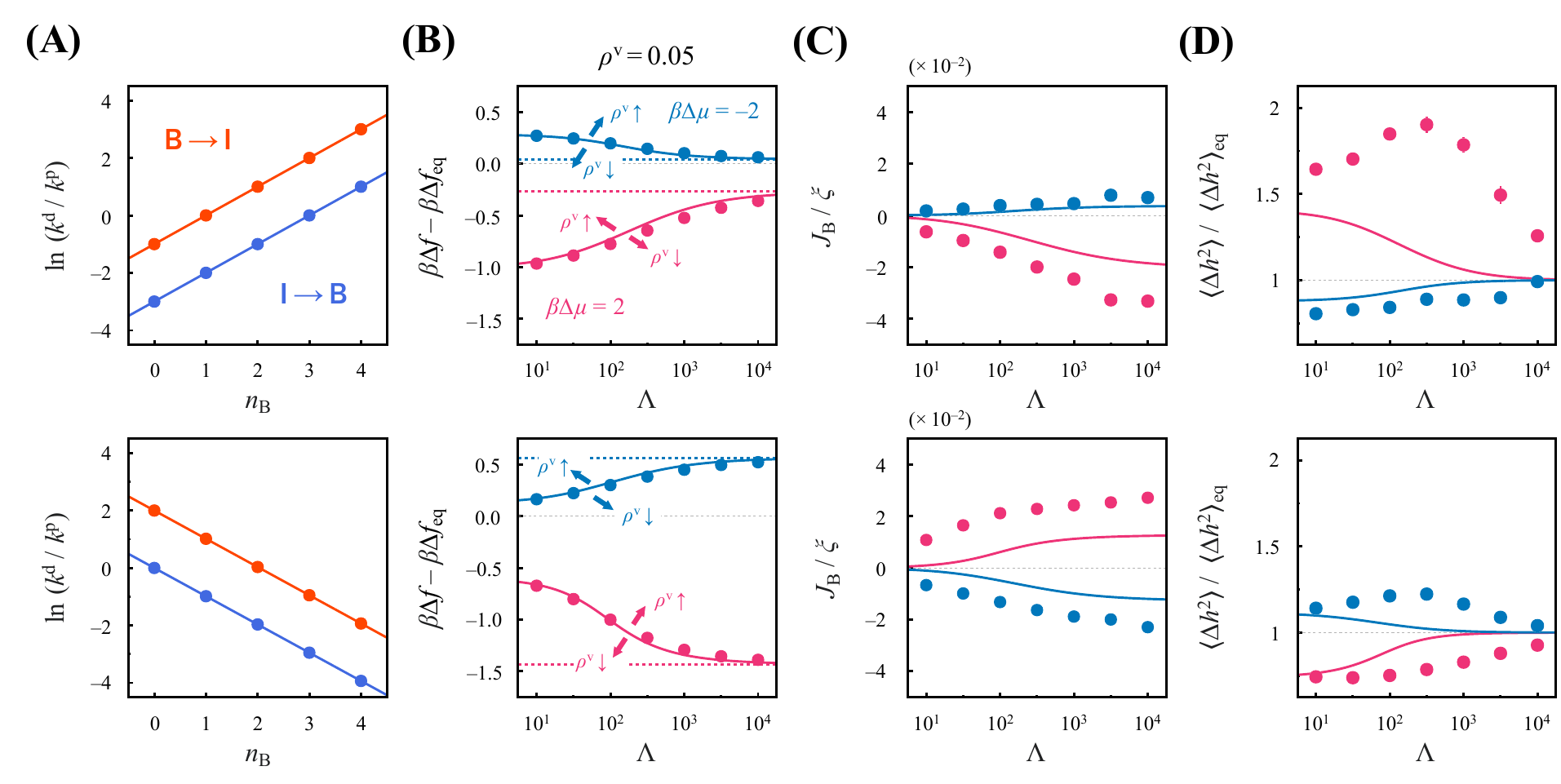}
    \caption{\textbf{Test of the nonequilibrium sharp-interface theory for alternative reaction schemes.}
    (A)~In these two schemes (upper and lower rows), the reaction rate constants in both the $\BI$ (orange) and $\IB$ (blue) directions depend on the local concentration of B-state molecules, $\nB$, but the ratio of the reaction rate constants between the driven (d) and passive (p) pathways, $k^\text{d}/k^\text{p}$, does not plateau (cf.~Fig.~1B in the main text).
    (B--D)~Simulations (marks) and predictions of the nonequilibrium sharp-interface theory (solid lines) for a macroscopically flat interface $(R\rightarrow\infty)$ are compared under far-from-equilibrium conditions at $\beta\Dmu = 2$ (pink) and $\beta\Dmu=-2$ (teal).
    (B)~Phase diagrams showing coexistence curves at a fixed far-field total concentration in the dilute phase, $\rhov = 0.05$.
    Moving away from the coexistence curves either increases $(\rhov\!\uparrow)$ or decreases $(\rhov\!\downarrow)$ the far-field concentration.
    The subscript ``eq'' indicates the value at equilibrium $(\Dmu = 0)$.
    Dotted lines show the predictions of the alternate sharp-interface theory that employs equilibrium interfacial boundary conditions.
    (C)~The average B-state diffusive flux density near the interface, $\JB/\xi$, and (D) the relative change in the mean-squared interfacial height fluctuations, $\langle \Dh^2 \rangle$.
    In panels C and D, all the data and predictions are made along the coexistence curves shown in panel B.
    In the upper row, the system parameters are $\beta\epsilon = -3.00$ and $\eta = 0.05$.
    In the lower row, the parameters are $\beta\epsilon = -2.95$ and $\eta = 1$, as in the main text.
    We note that the phase behavior in panels B--D is qualitatively similar to Fig.~4B--D in the main text up to a change in sign, indicating that the reaction scheme does not have to be strongly inhomogeneous on the interface (i.e., for $n_\text{B}\in\{1,2,3\}$) to induce nonequilibrium mesoscopic and interfacial effects.
    }
    \label{fig:SI_linear_kinetics}
\end{figure}

\begin{figure}
    \centering
    \includegraphics[width=0.6\linewidth]{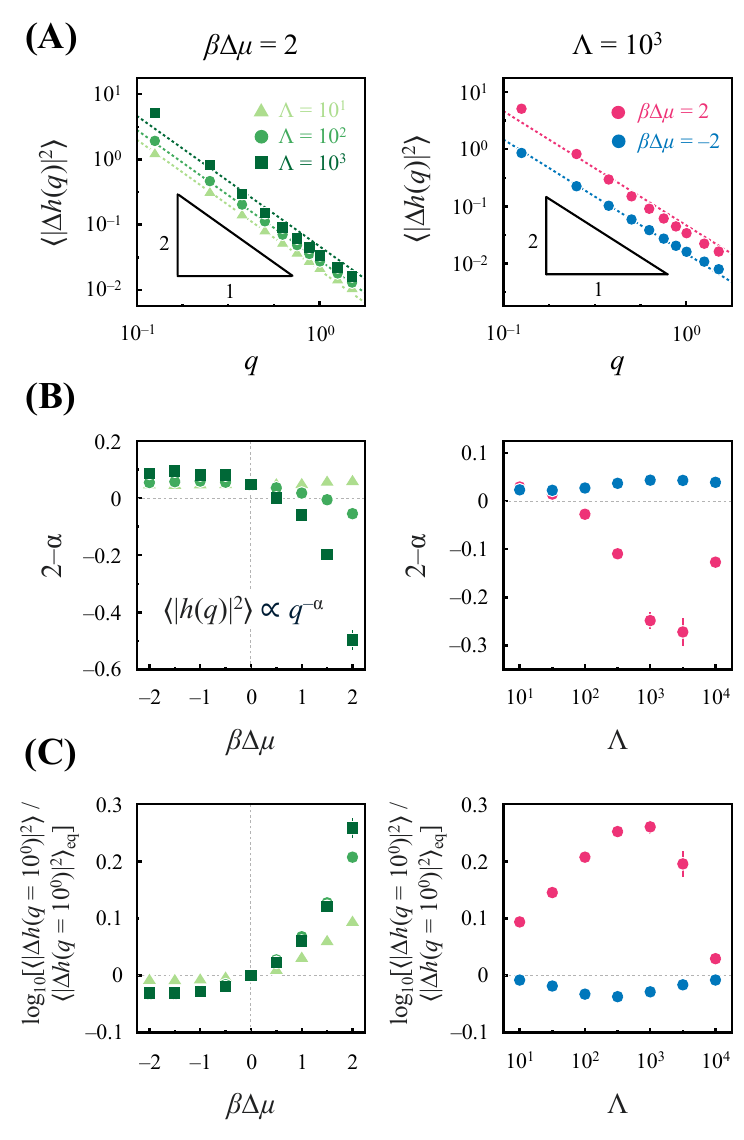}
    \caption{\textbf{Deviations from equilibrium capillary wave theory are apparent under far-from-equilibrium conditions.}
    Equilibrium capillary wave theory is tested for flat interfaces at nonequilibrium phase coexistence at constant $\Lambda$ (left column) and at constant $\beta\Dmu$ (right column) with the far-field total concentration in the dilute phase fixed to $\rhov = 0.05$.
    (A)~Example power spectra of interfacial height fluctuations, $\langle|\Dh(q)|^2\rangle$, with respect to a wavevector parallel to the interface, $q$.
    Example cases are shown at $\beta\Dmu = 2$ (left) and $\Lambda = 10^3$ (right).
    The dotted lines indicate fits to the equilibrium equipartition scaling, $\langle|\Dh(q)|^2\rangle \propto q^{-2}$~\cite{rowlinson2013molecular}.
    (B)~Deviations from the equipartition relation are quantified by the power-law exponent $\alpha$ that is fit to the simulated fluctuation spectrum by assuming the functional form $\langle |\Dh(q)|^2\rangle \propto q^{-\alpha}$ in the range $q\in[10^{-1},10^0]$.
    (C)~The overall vertical shift in the power spectrum relative to equilibrium (subscript ``eq'') is quantified by the change in the spectrum at $q = 10^0$.
    All data presented in this figure are obtained along the coexistence curve shown in Fig.~4A in the main text.}
    \label{fig:SI_capillary}
\end{figure}

\begin{figure}
    \centering
    \includegraphics[width=0.95\textwidth]{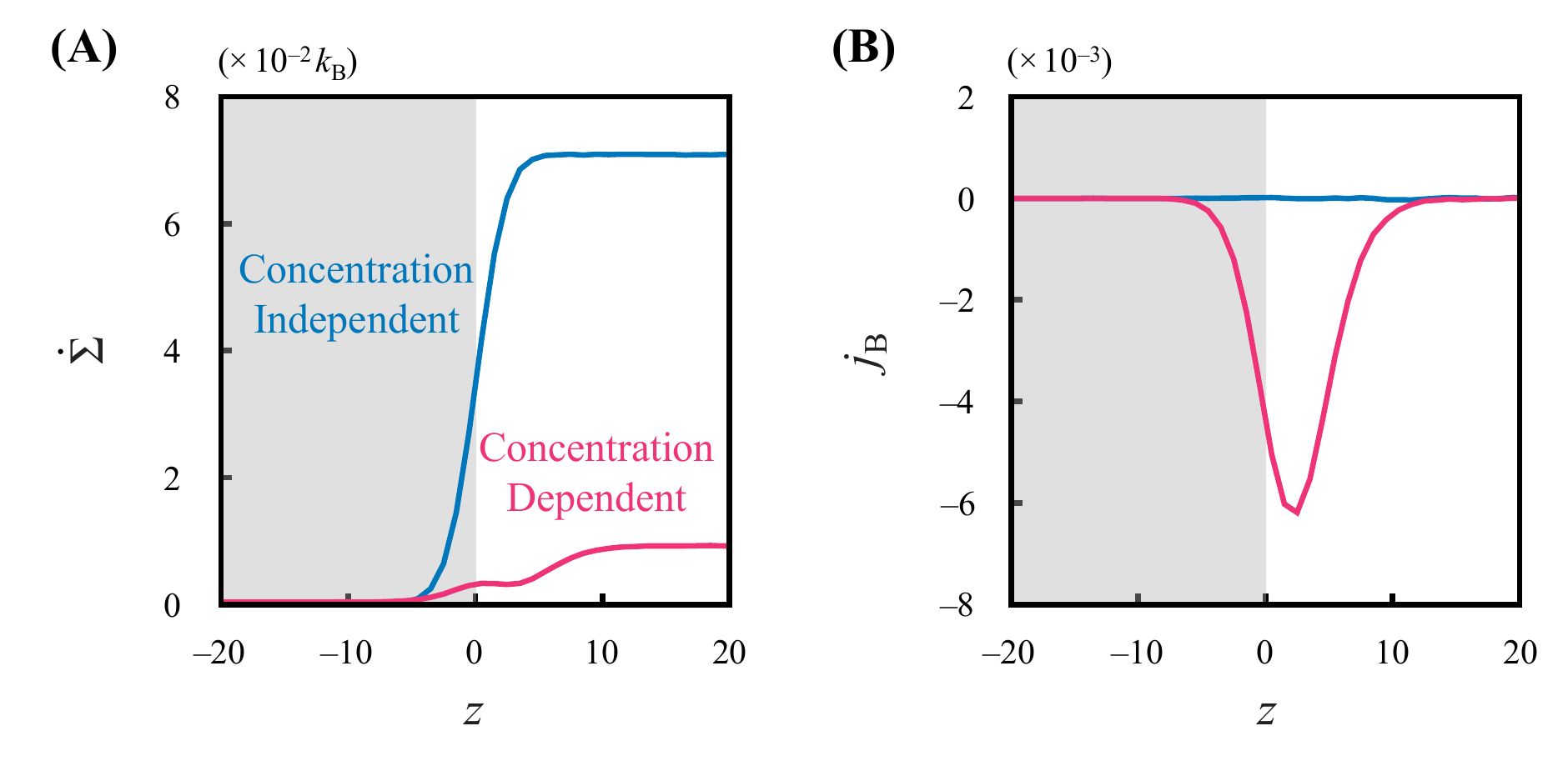}
    \caption{\textbf{Concentration-independent rate constants do not induce mesoscopic fluxes despite spatially inhomogeneous entropy production.}
    (A)~The entropy production rate density, $\dot{\Sigma}$, and (B)~the mesoscopic diffusive flux density for B-state molecules, $\jB$, assuming concentration-independent (teal) and concentration-dependent (pink) reaction schemes.
    The shaded and unshaded regions represent the coexisting condensed and dilute phases, respectively, separated by the Gibbs dividing surface.
    All data presented are obtained from flat-interface simulations with $\beta\Dmu = 2$, $\Lambda = 10^2$, and $\beta\Df = 1.7337$ for the concentration-dependent reaction scheme, and with $\beta\Df = 1.4024$ for the concentration-independent reaction scheme in order to match the far-field total concentration in the dilute phase, $\rhov = 0.05$.}
    \label{fig:SI_homogeneous}
\end{figure}

\begin{figure}
    \centering
    \includegraphics[width=\linewidth]{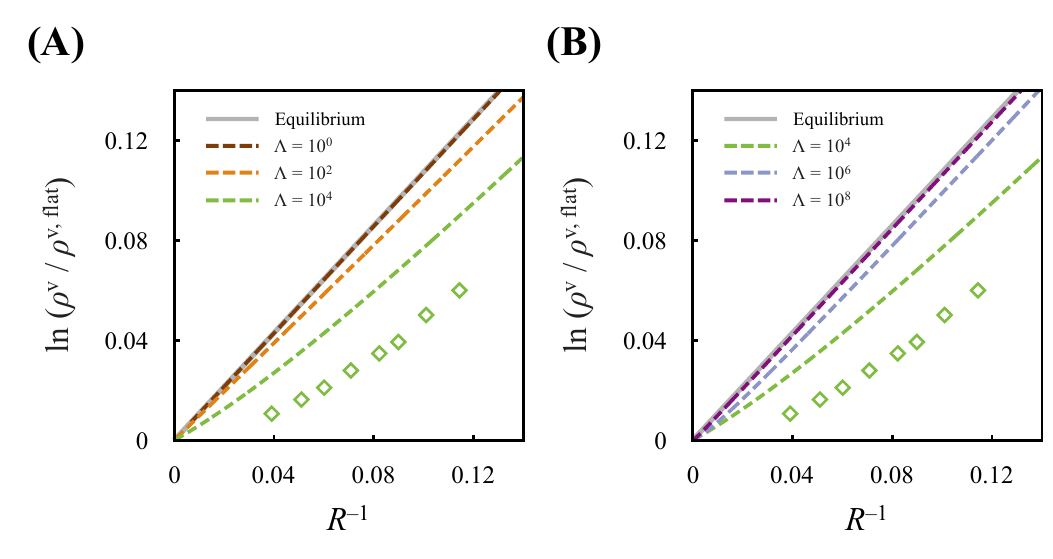}
    \caption{\textbf{The nonlinear effect of the diffusive flux on the size-scaling of finite-size droplets vanishes in both the reaction-dominant and diffusion-dominant limits.}
    Theoretical predictions for the nonequilibrium size-scaling relation are made at constant $\beta\Dmu = 2$ by following the predicted coexistence line at constant $\rhovf = 0.05$, as shown in Fig.~4A in the main text.
    This figure thus expands the range of $\Lambda$ values studied in Fig.~5B in the main text.  
    However, for demonstration purposes, the interfacial tension used in these predictions is fixed to the equilibrium value, $\beta\sigma_\text{eq} = 1.074$, when applying the boundary condition BC~(4) for curved interfaces.
    The values of $\Lambda$ considered are (A) $\Lambda = 10^0$ (dark brown), $10^2$ (light brown), and $10^4$ (light green), and (B) $\Lambda = 10^4$ (light green), $10^6$ (light blue), and $10^8$ (dark purple).
    The predictions at $\Lambda = 10^0$ in panel A and at $\Lambda = 10^8$ in panel B closely follow the equilibrium Young--Laplace scaling (gray solid lines), indicating that the nonlinearity vanishes in both the $\Lambda\rightarrow 0$ and $\Lambda\rightarrow \infty$ limits.
    In both panels, diamond marks indicate simulation measurements at $\Lambda = 10^4$; however, the light green dashed curves do not agree with the simulation data because the nonequilibrium interfacial tension, $\sigma \ne \sigma_{\text{eq}}$, is excluded from these calculations for demonstration purposes.}
    \label{fig:SI_sigma}
\end{figure}

\begin{figure}
    \centering
    \includegraphics[width=0.5\linewidth]{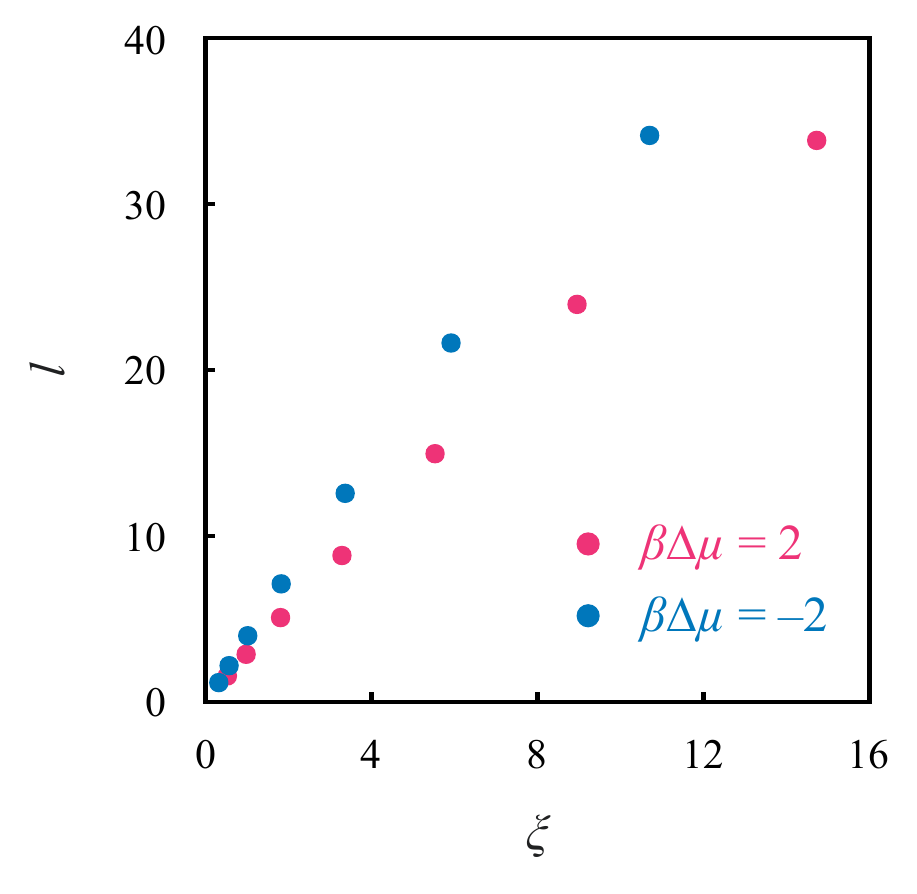}
    \caption{\textbf{Comparison of mesoscopic length scales in the dilute phase.}
    The reaction--diffusion length $\xi$ is defined according to \eqref{eq:xi}, whereas the correlation length $l$ is defined by \eqref{eq:correlation_length}.
    Marks indicate measurements of these two quantities along the coexistence curve at fixed $\beta\Dmu = 2$ (pink) and $-2$ (teal) with a fixed far-field total concentration in the dilute phase, $\rhov = 0.05$, as shown in Fig.~4A in the main text.
    The correlation length $l$ is measured in simulations of the dilute phase on a $300 \times 300$ lattice using $5\times 10^4$ event sweeps.
    The calculation of the reaction--diffusion length $\xi$ is described in \figref{fig:SI_RD_test}.
    }
    \label{fig:SI_xi_direct}
\end{figure}

\begin{figure} 
    \centering
    \includegraphics[width=0.75\textwidth]{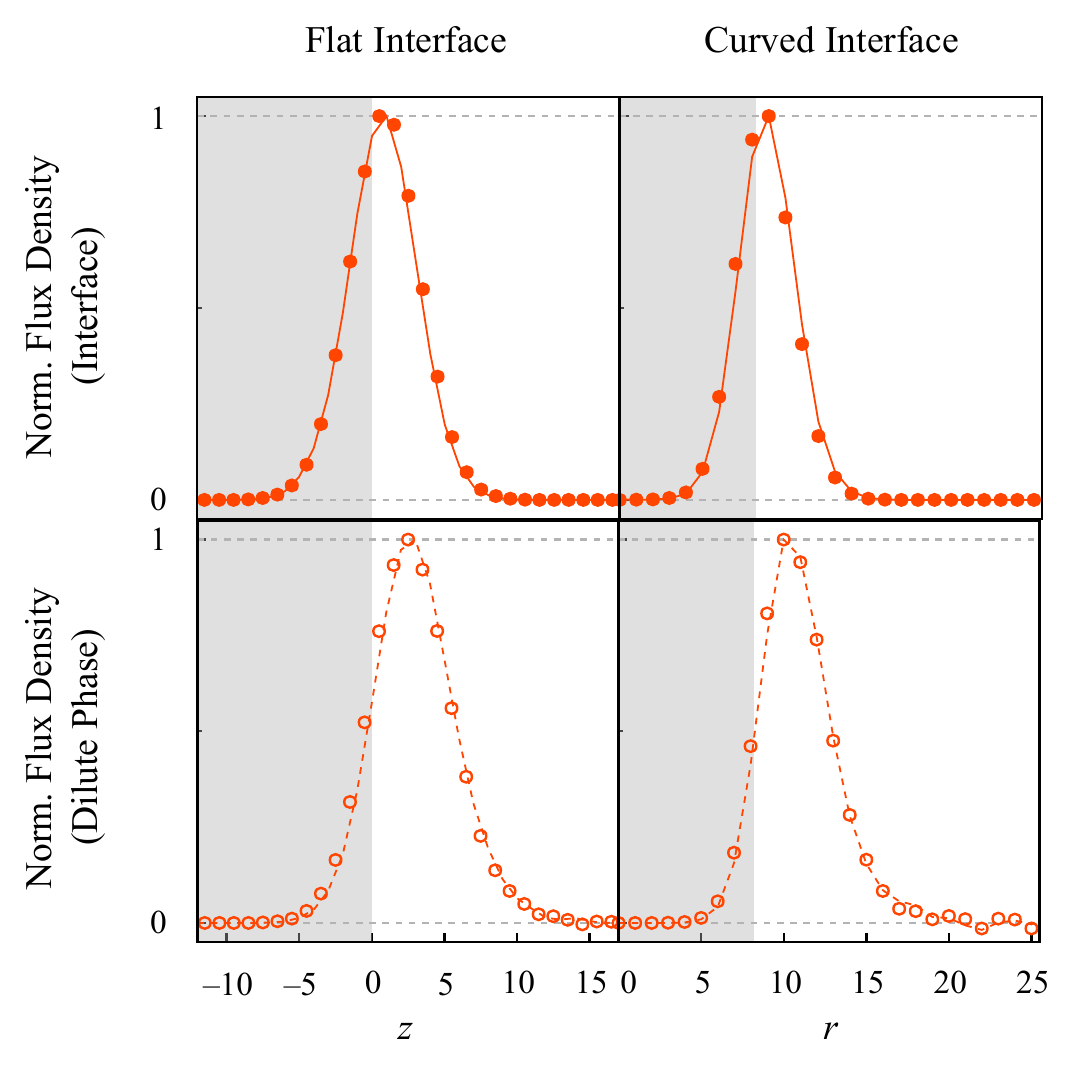}
    \caption{\textbf{The net steady-state reactive and diffusive fluxes are tightly correlated.} The interfacial (upper) and dilute-phase (lower) contributions to the net steady-state reactive flux densities, $\sBI-\sIB$ (marks), and diffusive flux densities, $\jB$ (lines), observed in the flat (left column) and curved (right column) interface simulations shown in Fig.~2B,C in the main text.
    The reactive and diffusive flux densities are normalized relative to their peak values.
    In all four panels, the shaded and unshaded regions indicate the condensed and dilute phases, respectively, separated by the Gibbs dividing surface.
    }
    \label{fig:SI_coupling}
\end{figure}

\begin{figure}
    \centering
    \includegraphics[width=0.5\linewidth]{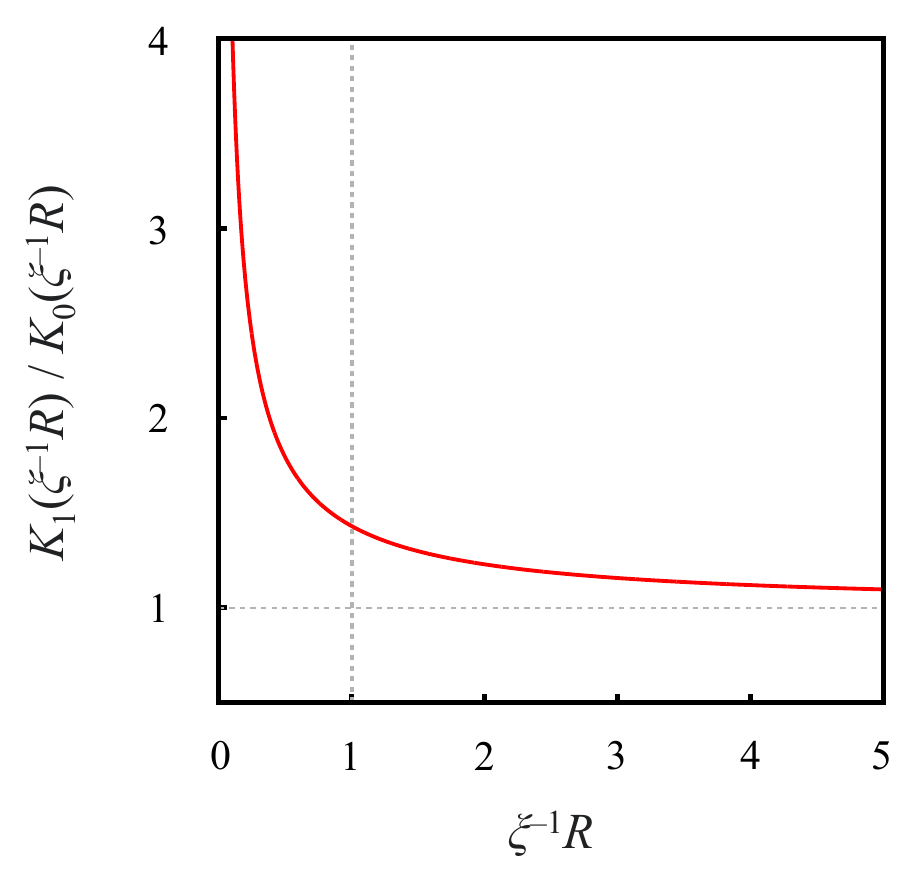}
    \caption{\textbf{Droplet-size dependence of the nonequilibrium diffusive flux correction in \eqref{eq:diffusive_flux_density_curved}.}
    The Bessel function term in this equation, $K_1 / K_0$, approaches unity in the limit $\xi^{-1}R \gg 1$, corresponding to the regime in which the reaction--diffusion length, $\xi$, is small compared to the droplet radius, $R$.
    However, $K_1 / K_0$ deviates substantially from unity when $\xi$ is comparable to or greater than the droplet size ($\xi^{-1}R \lesssim 1$).}
    \label{fig:SI_Bessel}
\end{figure}